\documentclass[%
 reprint,
nofootinbib,
amsmath,amssymb,
aps,
]{revtex4-2}

\usepackage{float}
\usepackage{graphicx}
\usepackage{dcolumn}
\usepackage{bm}
\usepackage{natbib}

\setcitestyle{authoryear,open={(},close={)}} 
\usepackage{url}
\usepackage{footnote}
\usepackage{graphicx}
\usepackage{txfonts}
\usepackage{tikz}
\usetikzlibrary{calc}
\usetikzlibrary{positioning}
\usetikzlibrary{shapes.geometric}

\usepackage{enumitem}

\newcommand{\aj}{Astron. J.} 
\newcommand{\aap}{Astron. Astrophys.} 
\newcommand{\aaps}{Astron. Astrophys. Suppl.} 
\newcommand{\mnras}{Mon. Not. R. Astron. Soc.} 

\begin{document}


\title{{\huge AutoSourceID-Classifier} \vspace{+1.5mm} \\
\LARGE Star-Galaxy Classification using a Convolutional Neural Network with Spatial Information}

\author{F.~Stoppa$^{1,6}$,
S.~Bhattacharyya$^{2}$,
R.~Ruiz de Austri$^{5}$,
P.~Vreeswijk$^{1}$,
S.~Caron$^{3,4}$,
G.~Zaharijas$^{2,9}$,
S.~Bloemen$^{1}$,
G.~Principe$^{12,13}$,
D.~Malyshev$^{14}$,
V.~Vodeb$^{2}$,
P.J.~Groot$^{1,10,11}$,
E.~Cator$^{6}$,
G.~Nelemans$^{1,7,8}$
}
\affiliation{
\begin{description}[labelsep=0.2em,align=right,labelwidth=0.7em,labelindent=0em,leftmargin=2em,noitemsep]
\item[$^{1}$] Department of Astrophysics/IMAPP, Radboud University, PO Box 9010, 6500 GL Nijmegen, The Netherlands
\item[$^{2}$] Center for Astrophysics and Cosmology, University of Nova Gorica, Vipavska 13, SI-5000 Nova Gorica, Slovenia
\item[$^{3}$] High Energy Physics/IMAPP, Radboud University, PO Box 9010, 6500 GL Nijmegen, The Netherlands
\item[$^{4}$] Nikhef, Science Park 105, 1098 XG Amsterdam, the Netherlands
\item[$^{5}$] Instituto de Física Corpuscular, IFIC-UV/CSIC, Valencia, Spain
\item[$^{6}$] Department of Mathematics/IMAPP, Radboud University, PO Box 9010, 6500 GL Nijmegen, The Netherlands
\item[$^{7}$] SRON, Netherlands Institute for Space Research, Sorbonnelaan 2, NL-3584 CA Utrecht, The Netherlands
\item[$^{8}$] Institute of Astronomy, KU Leuven, Celestijnenlaan 200D, B-3001 Leuven, Belgium
\item[$^{9}$] Institute for Fundamental Physics of the Universe, Via Beirut 2, 34151 Trieste, Italy
\item[$^{10}$] Department of Astronomy, University of Cape Town, Private Bag X3, Rondebosch, 7701, South Africa
\item[$^{11}$] South African Astronomical Observatory, P.O. Box 9, Observatory, 7935, South Africa
\item[$^{12}$] Dipartimento di Fisica, Universit\'a di Trieste, I-34127 Trieste, Italy
\item[$^{13}$] Istituto Nazionale di Fisica Nucleare, Sezione di Trieste, I-34127 Trieste, Italy
\item[$^{14}$] Erlangen Centre for Astroparticle Physics, Nikolaus-Fiebiger-Str. 2, Erlangen 91058, Germany
\end{description}

}%

\date{\today\\}

\begin{abstract}

\noindent
\textit{Aims.} {Traditional star-galaxy classification techniques often rely on feature estimation from catalogues, a process susceptible to introducing inaccuracies, thereby potentially jeopardizing the classification's reliability. Certain galaxies, especially those not manifesting as extended sources, can be misclassified when their shape parameters and flux solely drive the inference. We aim to create a robust and accurate classification network for identifying stars and galaxies directly from astronomical images. By leveraging convolutional neural networks (CNN) and additional information about the source position, we aim to accurately classify all stars and galaxies within a survey, particularly those with a signal-to-noise ratio (S/N) near the detection limit.}

\noindent
\textit{Methods.} {The AutoSourceID-Classifier (ASID-C) algorithm developed here uses 32x32 pixel single filter band source cutouts generated by the previously developed ASID-L code. ASID-C utilizes CNNs to distinguish these cutouts into stars or galaxies, leveraging their strong feature-learning capabilities. Subsequently, we employ a modified Platt Scaling calibration for the output of the CNN. This technique ensures that the derived probabilities are effectively calibrated, delivering precise and reliable results.}

\noindent
\textit{Results.} {We show that ASID-C, trained on MeerLICHT telescope images and using the Dark Energy Camera Legacy Survey (DECaLS) morphological classification, outperforms similar codes like SourceExtractor. ASID-C opens up new possibilities for accurate celestial object classification, especially for sources with a S/N near the detection limit. Potential applications of ASID-C, like real-time star-galaxy classification and transient's host identification, promise significant contributions to astronomical research.}

\end{abstract}


\maketitle


\section{Introduction}

Current and future large-scale astronomical photometric surveys are amassing, and will continue to amass, vast quantities of photometric data, including millions of images containing billions of stars and galaxies. This data influx necessitates a series of processing steps, such as source localization, anomaly detection, feature extraction, and star-galaxy classification, to be further optimized for efficiency in terms of both time and computational resources. With the advent of CMOS detectors capable of capturing large-format images of the night sky at cadences exceeding 1Hz, traditional methods and software for data processing will be inadequate to keep pace with the data acquisition rate.

To address this challenge, our research aims to enhance and streamline existing methods using machine learning tools. In AstroSourceID-Light (ASID-L, \citealp{Stoppa2022}), we demonstrated the potential for rapid and accurate source identification on images, achieving this in a fraction of the time required by currently used methods. Building on this foundation, we now focus on star-galaxy classification, a fundamental data-processing task and often the initial step for the scientific exploitation of survey data. Despite advances in source localization and feature extraction algorithms, there remains significant potential for improvement in star-galaxy classification methods, which are often applied from catalogues rather than directly from source images \citep{Weir1995, Ball2006, Vasconcellos2011, Sevilla-Noarbe2018}. Current classification methods, including classification tree methods based on the morphological features of the sources and Bayesian approaches that integrate available source information with prior knowledge about nearby star and galaxy populations, are prevalent but have room for enhancement \citep{Henrion2011, Lopez2019}.

Like many areas of astronomy, data analysis in star-galaxy classification has seen a surge in machine learning applications. The first machine learning application for star-galaxy classification was introduced in \cite{Odewahn1992}, and it quickly became a core component of the astronomical image processing software SourceExtractor \citep{Bertin1996}. More recently, a series of studies have employed various neural network architectures to address the star-galaxy classification problem \citep{Odewahn2004, Fadely2012, Cabayol2018}, with significant progress made in \cite{Kim2016}. In this study, the authors demonstrated the effectiveness of convolutional neural networks (CNNs) in learning features directly from multi-band ($ugriz$) optical images of different sources. When applied to the CFHTLenS dataset \citep{Heymans2012}, their network was able to generate a classification score that matched the accuracy of a random forest-based algorithm \citep{breiman2017classification}, but with better-calibrated probability estimates, indicating a promising avenue for future research in this field.

\begin{figure}[ht]
\centering
\includegraphics[scale=.32]{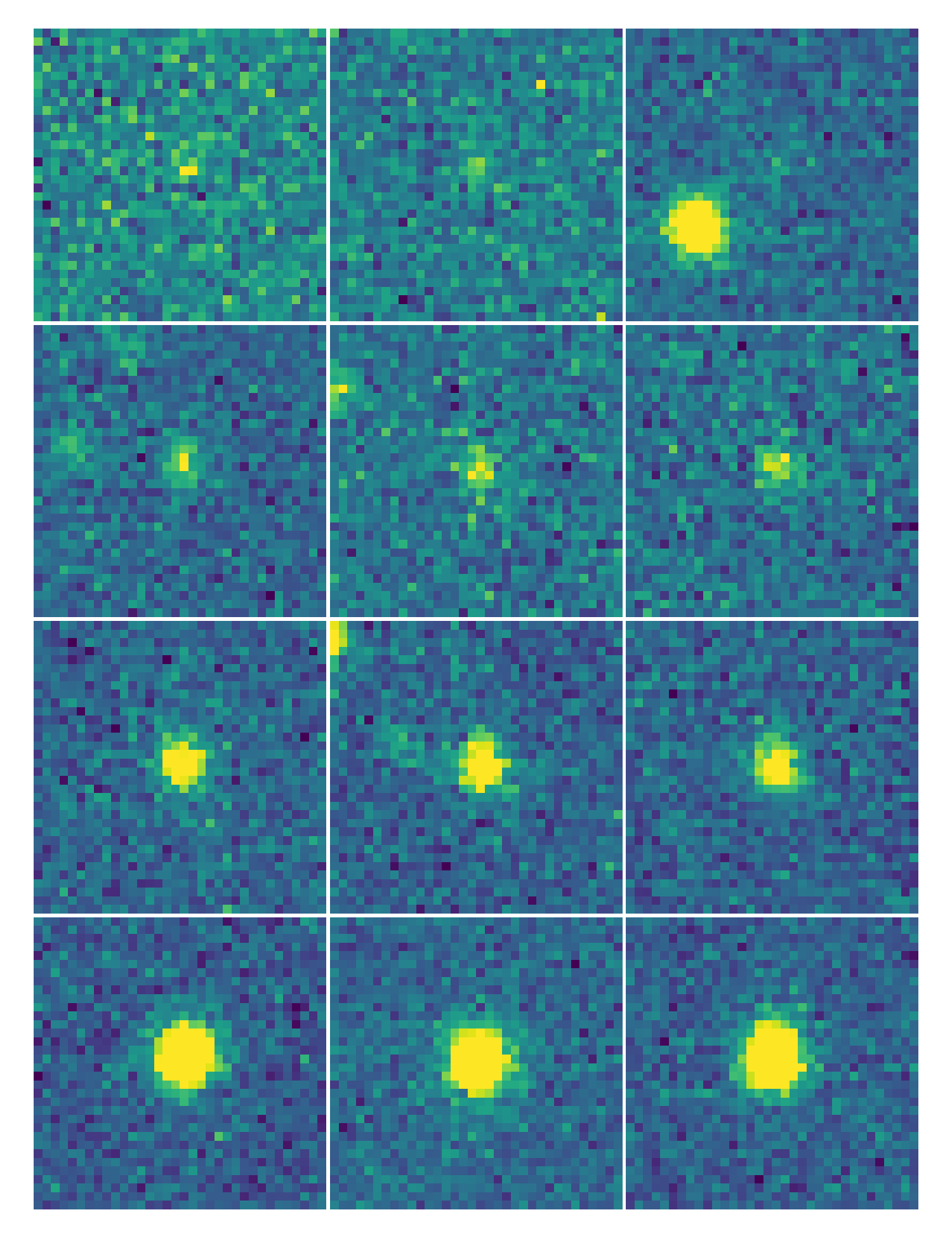}
\includegraphics[scale=.32]{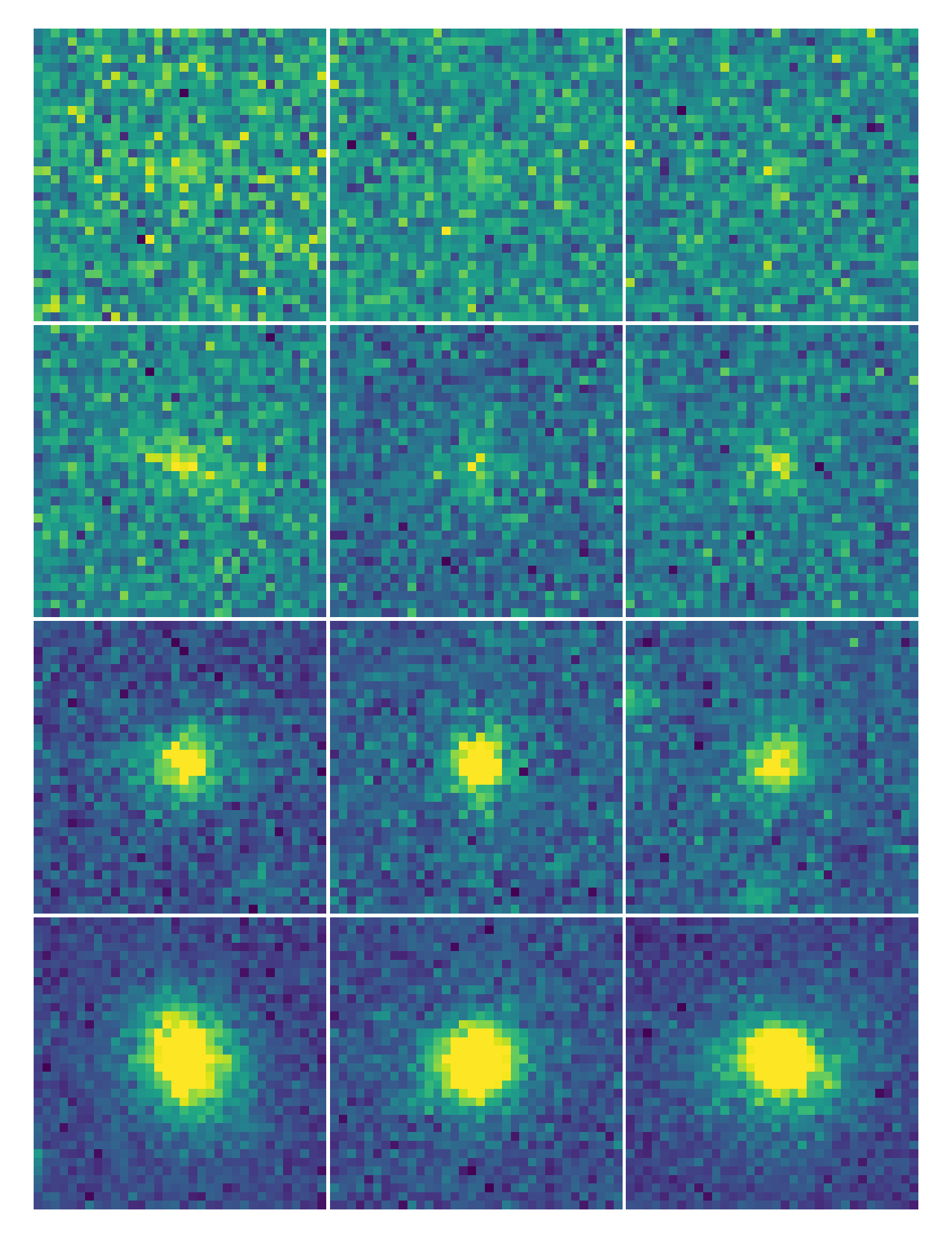}
\caption{Cutouts of stars (left) and galaxies (right) for S/N = \{4, 10, 25, 65\} (top to bottom). The source is always at the centre of the cutout. At low S/N, sources are barely discernible, making their classification a complex task.
}
\label{fig: cutouts}
\end{figure}

Most of the methods mentioned above operate under the assumption that galaxies manifest as extended sources while stars appear as point sources. However, this distinction becomes less clear at fainter magnitudes and higher redshifts, where the morphological features of galaxies are less discernible. As a result, the effectiveness of these methods decreases under such conditions. Our study addresses this challenge by aiming to accurately classify all stars and galaxies within a survey, with a particular focus on those with a signal-to-noise ratio (S/N) near the detection limit, as illustrated in Fig. \ref{fig: cutouts}.

This task presents a unique set of challenges, primarily due to the need to identify a limited range of morphological characteristics. Our deliberate choice of single-band images further compounds these challenges. Although these single-band images inherently contain less information than their multi-band counterparts, they offer a significant advantage in terms of data acquisition simplicity. Eliminating the need for repeated observations of the same source makes them a more practical choice, particularly for small to medium-sized optical telescopes when multi-band data may not always be available. This characteristic of single-band imagery is not a constraint but rather an opportunity to refine our classification methodologies.

Given these challenges and the potential of single-band imagery, we have developed the AutoSourceID-Classifier (ASID-C). This tool is designed to take optical image cutouts of sources, which in our case are retrieved with ASID-L, and their positions in the full telescope image as input, and then output a probability for each source being either a star or a galaxy. To ensure the validity of our classification model's predictions, we have employed an enhanced Platt scaling method \citep{platt1999} to calibrate the network's outputs.\\
The method presented here\footnote{\url{https://github.com/FiorenSt/AutoSourceID-Classifier}} is the fourth deep learning algorithm developed in the context of MeerLICHT/BlackGEM telescopes \citep{Bloemen2016,groot2022}, following MeerCRAB, an algorithm for classifying real and bogus transients \citep{Hosenie2021}, ASID-L for source localization \citep{Stoppa2022}, and ASID-FE for feature extraction \citep{Stoppa2023}. Building on these deep learning algorithms, our ultimate goal is to establish the first fully automated machine learning detection pipeline for small to medium-sized optical telescopes, thereby facilitating more efficient and accurate astronomical observations and analyses.

The rest of this paper is organized as follows: Section \ref{sec: Data} provides details on the datasets used and the preprocessing steps undertaken. Section \ref{sec:Method} describes the model, while Section \ref{sec: Results} presents the results, including a comparison with SourceExtractor \citep{Bertin1996}, a widely used tool for optical image analysis. Finally, in Section \ref{sec: Conclusion and Discussion}, we present our conclusions and discuss potential scientific applications of our tool.

\section{Data}
\label{sec: Data}

To develop our classifier, we constructed a dataset comprising source cutouts from images captured by the MeerLICHT telescope, paired with morphological classifications of sources from the Dark Energy Camera Legacy Survey (DECaLS, \citealp{Blum2016}). To the best of our knowledge, this dataset, containing approximately 12 million source cutouts, is the largest ever assembled for star-galaxy image classification in a machine learning context. In the following section, we outline the telescopes and catalogues used to create this dataset and provide a brief overview of the preprocessing steps involved in obtaining the image cutouts.

\subsection{MeerLICHT images}

The MeerLICHT telescope, which pairs a 65cm diameter primary mirror with a 10.5k x 10.5k CCD detector, creates a $2.7$ square degree monolithic field-of-view, sampled at $0.56"/pix$ \citep{Bloemen2016}. With an average image quality of $2-3"$, point sources are sampled at $4-6$ pixels per FWHM. The telescope's primary function is to enable simultaneous transient detection at radio and optical wavelengths, working in conjunction with the MeerKAT radio telescope \citep{Jonas2016}. The available filter set includes the SDSS $ugriz$ filters and a broader $q$-band filter ($440-720nm$), which roughly combines SDSS $g+r$. Images captured by the telescope undergo processing at the IDIA/ilifu facility through the BlackBOX software\footnote{\url{https://github.com/pmvreeswijk/BlackBOX}} (Vreeswijk et al., in prep), which handles standard image processing tasks such as source detection, astrometric and photometric calibration, creation of the position-dependent point spread function (PSF), image subtraction, photometry, and transient detection.

\begin{figure}[ht]
\centering
\includegraphics[scale=.35]{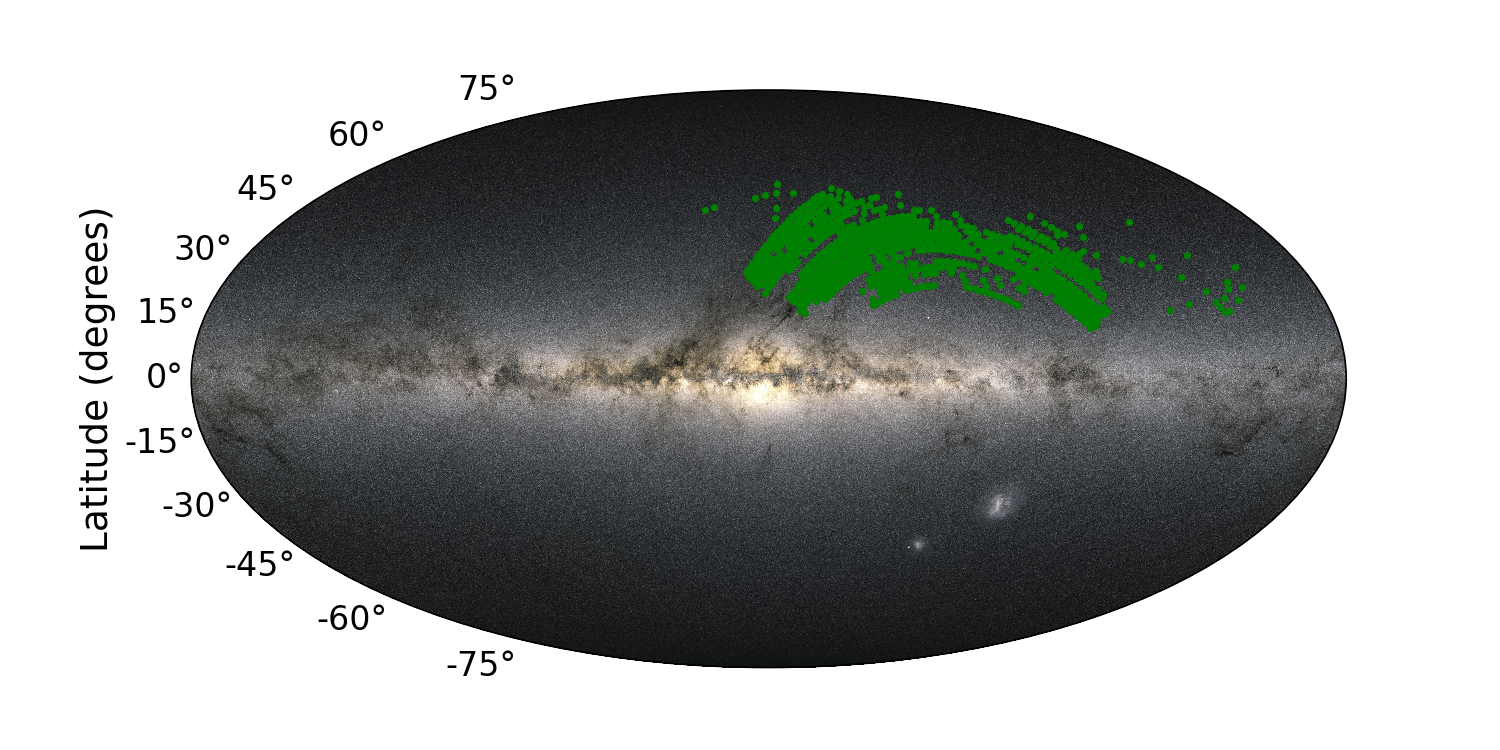}
\caption{Sky coordinates of the 718 MeerLICHT images (green dots) overlayed on an image of the Galactic Plane. This visualization demonstrates the diverse array of spatial densities covered by our dataset, facilitating a thorough evaluation of our model's capacity to classify regions with varying degrees of crowdedness.}
\label{fig: sky_coord}
\end{figure}

To ensure diversity, we generated the 12 million source cutouts from 718 non-overlapping, full-field images captured by the MeerLICHT telescope across various filters. These images were marked ``green'' in the pipeline, indicating they were free from obvious anomalies. However, we did not impose any additional restrictions based on visibility parameters. As depicted in Fig. \ref{fig: sky_coord}, the full-field images cover a wide range of spatial densities, from densely populated regions along the Galactic Plane to areas characterized by a sparse distribution of sources. This comprehensive coverage allows us to train and test the neural network's ability to classify regions of varying crowdedness.

For each source in the images, a cutout of 34x34 pixels is created with the source, previously localized with ASID-L, at its centre. This size was chosen to ensure that the vast majority of sources are comfortably enclosed within the cutout boundaries. For galaxies larger than 19 arcseconds, the cutout will be smaller than the source itself; however, this will not prevent our network from learning that such images are likely to be galaxies. In addition, we match these localized sources with their counterparts in MeerLICHT's catalogues. These catalogues, produced using SourceExtractor \citep{Bertin1996}, provide approximate estimates for flux, location, and a stellarity parameter, which are useful for additional testing on the results of ASID-C. Finally, out of these, we retain the 12 million cutouts that have a matching source in the DECaLS dataset, which we will discuss in more detail in the next section.

\subsection{Legacy survey}

To complement the MeerLICHT images and provide a robust basis for star-galaxy classification, we utilized the catalogues from the Dark Energy Camera Legacy Survey (DECaLS), part of the DESI Legacy Imaging Surveys. \\
The DESI Legacy Imaging Surveys aim to map 14,000 square degrees of extragalactic sky in three optical bands (g, r, z) and combine it with four mid-infrared bands from the Wide-field Infrared Survey Explorer (WISE) \citep{Dey2019, Schlegel2021}. This ambitious project is accomplished through three imaging projects that comprise the Legacy Surveys: DECaLS, the Beijing-Arizona Sky Survey (BASS, \citealp{Zou2019}), and the Mayall z-band Legacy Survey (MzLS, \citealp{Silva2016}).
The current data release, Data Release 10 (DR10), is the tenth public data release from these surveys. Source detection in DR10 is performed using a filter matched to the point spread function (PSF) and spectral energy distribution (SED) of the sources on the stacked images, featuring a $6\sigma$ detection limit. Each image is processed using its PSF model to detect sources, and these processed images are then combined in a weighted manner to improve the detection of point sources. DR10 provides a classification of sources into five types based on their shape and structure, determined using a multi-band, multi-epoch photometric model, including one for point sources and four for galaxies: round exponential galaxies ("REX"), deVaucouleurs ("DEV") profiles, exponential ("EXP") profiles, and Sersic ("SER") profiles.

DECaLS, in particular, offers significant overlap with the fields observed by MeerLICHT, making it an ideal choice for our study. The depth and quality of the imaging data provided by DECaLS allow for reliable and detailed morphological information, which is crucial for differentiating between stars and galaxies, especially at faint magnitudes where the morphological features of galaxies may be less discernible.
For our dataset, we only use DR10 sources detected by both ASID-L and the MeerLICHT pipeline in the 718 MeerLICHT images, as this allows us to use additional features estimated by the MeerLICHT pipeline, such as flux and S/N.

\begin{figure}[ht]
\centering
\includegraphics[scale=.34]{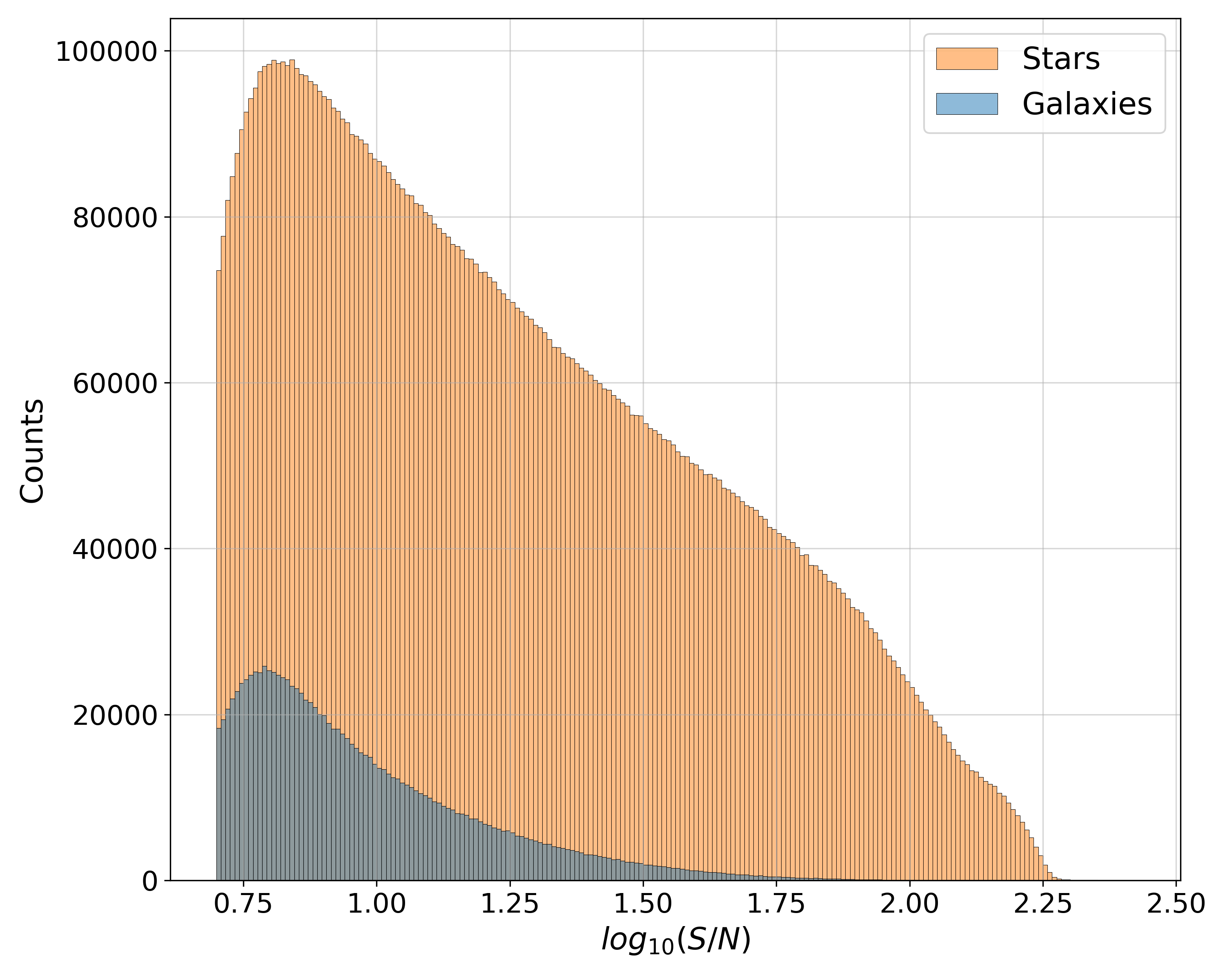}
\caption{Source counts by morphology as a function of S/N for our dataset. It reveals a $10:1$ disproportion favouring stars and a distinct difference in S/N, where stars consistently display higher values.}
\label{fig: sn_distribution}
\end{figure}

As depicted in Fig. \ref{fig: sn_distribution}, the resulting dataset from the match of MeerLICHT and DECaLS sources is characterized by an imbalance, with a 10:1 ratio favouring stars over galaxies and a noticeable difference in the S/N range. This imbalance underscores the challenge of our task; while the source population is predominantly stars at higher S/N values, the task of distinguishing between stars and galaxies becomes increasingly complex at lower S/N values.

\subsection{Crossmatching with additional Galaxies and Quasar's catalogues}

To evaluate the accuracy of the galaxy labels within our dataset, we performed a crossmatch operation with external catalogues. For this crossmatch, we incorporated galaxies from the Two Micron All Sky Survey (2MASS, \citealp{Skrutskie2006}) and the Lyon-Meudon Extragalactic Database (LEDA, \citealp{Paturel1995}). These galaxies were then cross-referenced with all the 12 million sources in our dataset, preserving those within a proximity of less than 0.56 arcsecond (approximately 1 pixel). Of all the matching sources in the refined subset, 98.5\% were accurately identified as galaxies within the Legacy survey dataset. The remaining 1.5\% misclassified sources pose a minor concern for our analysis, likely attributable to chance alignment and the fact that the sources we use are in a sky section first incorporated in the Legacy Survey in the latest DR10 release.

We further cross-referenced our catalogue with a known list of quasars retrieved from the SIMBAD astronomical database \citep{Wenger2000}. Quasars, or quasi-stellar objects, are exceptionally luminous active galactic nuclei fueled by the accretion of material onto supermassive black holes at the heart of distant galaxies. Within our star-galaxy classification dataset, a minuscule fraction of the images, approximately $0.006\%$, are quasars. Due to their stellar-like appearance and intense luminosity, quasars are frequently misclassified as stars, a trend evident in our dataset where $95\%$ of these quasars are erroneously labelled stars.

\subsection{Data Split, Augmentation and Calibration}
\label{sec: Data Split and Augmentation}

In this section, we discuss the process of data splitting, calibration, and augmentation, which are crucial steps in preparing our dataset for effective machine learning model training and evaluation.

The image cutouts dataset is divided into four subsets: $50\%$ for training, $20\%$ for validation, $20\%$ for testing, and $10\%$ for calibration. The training, validation, and test sets are used to train the model and evaluate its performance, while the calibration set is used to fine-tune the classifier's probability predictions. 
It's important to note that the split is made considering the disproportion of stars with respect to galaxies in the dataset, which is approximately 10:1. This ensures that the distribution of stars and galaxies in each subset reflects the actual distribution in the dataset, which is crucial for training a model that can accurately classify these celestial objects.

The dataset consists of 34x34 pixel cutouts, specifically designed to enable an augmentation step during the training stage aimed at improving the prediction of the network. We enhance the model's robustness and generalizability by applying a random one-pixel shift in any cardinal direction (up, down, left, or right) to each image. This random one-pixel shift reduces the cutout's dimensions to 32x32 pixels, contributing to a more diverse training dataset. Since the shift is only one pixel, it does not substantially impact the source's information; however, it introduces variability in the training set that can be leveraged to identify sources in different positions. Additionally, this shifting helps mimic the potential misalignment of the sources in the cutouts in real scenarios.

Incorporating the calibration set ensures that ASID-C produces well-calibrated probability predictions, ultimately improving classification performance. The calibration process, described in Sec. \ref{sec: platt scaling}, adjusts the classifier's output probabilities to better align with the true class probabilities, reducing the potential for over or underestimating probabilities. This step enhances the reliability and usefulness of the probability predictions, particularly in applications where accurate probability estimates are essential for downstream analysis or decision-making.

\section{Method}\label{sec:Method}

In the field of image classification, machine learning techniques have been instrumental in transforming the way we analyze and interpret data. Among these techniques, Convolutional Neural Networks (CNNs) have emerged as a powerful tool due to their ability to process image data in a robust and flexible manner. Introduced by \cite{Fukushima1982} and \cite{lecun1995}, CNNs have found widespread use in the computer vision community, providing a robust foundation for our work.

In this section, we outline our specific implementation of a machine learning model for star-galaxy classification. This model is designed to tackle the intricacies of our dataset, providing an optimized and custom-tailored approach to meet this classification challenge. Although the model is specifically tailored for our dataset and the star-galaxy classification task, its general structure and approach can be adapted for other astronomical datasets, telescopes, and purposes involving image data. This flexibility makes our model a versatile tool that can contribute to a wide range of astronomical research tasks.

\subsection{Model}
\label{subsec:Model}

Our model utilizes a Convolutional Neural Network (CNN), a type of neural network particularly effective for image analysis. In a CNN, an image is processed through multiple layers to generate feature maps. These maps are created by convolving each input feature map with a set of weights known as filters. Each feature map utilizes a distinct set of filters, allowing for a rich representation of the input data. This mechanism is at the core of our model's ability to analyze and classify the celestial images in our dataset.

While previous studies such as \cite{Kim2016} have utilized CNN architectures for star-galaxy classification, they often rely on multiple filter band images to enhance the classification results. In contrast, our approach simplifies this requirement by using single-band images. This makes our model versatile and independent of the specific filter applied.

Our deep neural network is designed with a dual-branch structure, specifically crafted to process image data and spatial information as inputs. This architectural choice is informed by the unique challenges of classifying celestial objects in optical telescope images, specifically, in our case, those from the MeerLICHT telescope.
Spatial information serves a crucial role in our model, aiding the classifier in handling variations in the appearance of sources based on their location within the full-field MeerLICHT image. The Point Spread Function (PSF) of sources, which influences their appearance, varies depending on their position within the image. For instance, sources at the centre of the image are symmetrical, while those near the edges exhibit asymmetry. Furthermore, sources farther from the image's centre display an elongated PSF, causing stars to resemble galaxies.
By integrating spatial information into our model, we can effectively account for these variations, thereby enhancing the accuracy of our model in classifying stars and galaxies. We validated the effectiveness of this approach by comparing the model’s performance with and without the spatial information branch. The results are shown in Appendix \ref{sec: appendix_I}.

\begin{figure*}[ht]
\centering
\includegraphics[scale=.48]{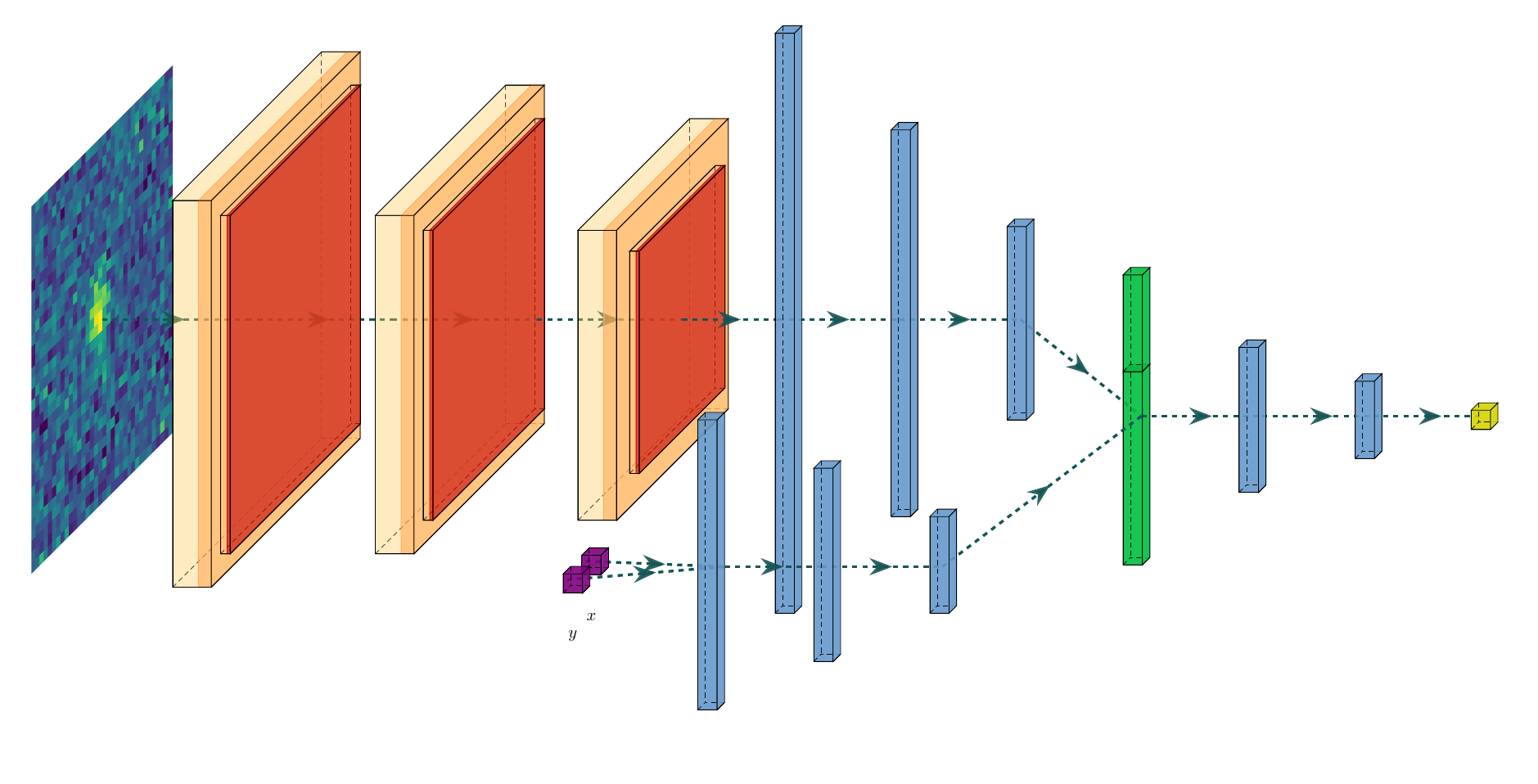}
\caption{The dual-branch network architecture used in this study. The primary branch processes the images through a series of convolutional layers (yellow), max-pooling layers (red), and dense layers (blue) to extract key features. The secondary branch processes the spatial information (pixel coordinates $[x, y]$) through multiple dense layers. Both branches merge through a concatenation layer (green), followed by additional dense layers and a final softmax layer for classification. This architecture allows the model to leverage image features and spatial information for more accurate star-galaxy classification.}
\label{fig: architecture}
\end{figure*}

As shown in Fig. \ref{fig: architecture}, the model is a dual-branch neural network that accepts image data and spatial information as inputs. The imaging branch, a CNN, processes 32x32 pixel images of celestial objects, while the spatial branch handles the 2-dimensional location data. The CNN comprises three convolutional layers with 32 filters each, followed by max-pooling layers. The spatial branch, a fully connected network, includes two dense layers with 64 and 32 neurons, respectively. Outputs from both branches are concatenated and passed through two additional dense layers before reaching the final sigmoid activation function, which generates a value between 0 (galaxy) and 1 (star).

\noindent
We train the model using the Adam optimizer \citep{AdamOpt}. We implement several strategies to achieve better accuracy and convergence and prevent overfitting. As the network gets trained over multiple iterations and approaches the minima of the loss landscape, it is typically suggested to have a lower learning rate $(\eta _t)$ for improving convergence \citep{LRdecay}. In our approach, we implement exponential decay with an initial learning rate $(\eta _0)$ of $0.001$ that decays exponentially after eight epochs (iterations) with a decay rate $\gamma = 0.99$.

\noindent
To prevent overfitting, we implement early stopping, which ensures the training stops if the loss for the validation set does not improve over ten epochs. The model's performance is assessed using precision, recall, and the area under the receiver operating characteristic (ROC) and precision-recall (PR) curves. Our approach allows for a low number of parameters, approximately 72k, making the algorithm fast and suitable for the needs of an automatic detection pipeline. 

We use TensorFlow\footnote{https://www.tensorflow.org/}\citep{tf:whitepaper} to implement the model and evaluate all possible hyperparameters with Weights \& Biases\footnote{https://wandb.ai/site}, a machine learning platform for developers to track, version control, and visualize results, which is especially helpful in our case for fixing hyperparameters. The results presented below were computed using an NVIDIA GeForce RTX 2080 GPU.

\subsection{Loss function}\label{sec:lossf}

In this work, we employ the Binary Focal Loss as our loss function, initially introduced by \cite{Lin2017} to tackle the class imbalance issue in object detection tasks. The Focal loss is designed to prioritize hard-to-classify examples while reducing the weight of easy examples. It is defined as:

\begin{equation}
FL\,(y, p) = -\alpha_t \, (1 - p_t)^{\gamma} \log(p_t) \ , \
\label{eq: loss}
\end{equation}

\noindent
where

\begin{equation}
\alpha_t, \, p_t =
\begin{cases}
\alpha, \; p; & \, y=1 \\
1-\alpha, \; 1-p; & \, y=0 \\
\end{cases}
\end{equation}

Here, $y$ is the true label, $p$ is the predicted probability, $\alpha$ is a weighting factor to balance the two classes, and $\gamma$ is a focusing parameter that adjusts the rate at which easy examples are down-weighted.
When $\alpha=1$ and $\gamma=0$, the Binary Focal Loss simplifies to the well-known Binary Cross Entropy loss \citep{Cox1958}. The Binary Cross Entropy loss is a common function for binary classification tasks, but it can be less effective when there is a class imbalance. 

While the Focal Loss was initially designed for one-stage detectors in computer vision tasks with a significant class imbalance between foreground and background (on the order of 1000:1), we adapt this loss function to address the class imbalance between stars and galaxies (approximately 10:1) in our dataset. 
However, as shown in Fig. \ref{fig: Calibration}, we found no substantial difference between the results of any of the focal loss models, especially after the calibration method introduced in the next section. Therefore, a focal loss with parameters $\alpha=1$ and $\gamma=0$, which equates to a Binary Cross Entropy, emerges as the optimal choice.

The lack of improvement from the Focal loss, despite the 10:1 imbalance, can be attributed to the concept of the Effective Sample Size (ESS, \citealp{BARTOSZEK2016,cui2019}). The ESS of our dataset, which is a measure of the number of independent observations that a given dataset is equivalent to, is large enough to allow for accurate learning despite the imbalance. This suggests that when the dataset is sufficiently large, the use of the focal loss may not bring any additional benefit.\\
Although the focal loss does not yield a substantial improvement over the standard cross-entropy loss in our specific case, it offers an opportunity to investigate the impact of varying the focusing parameter $(\gamma)$ and the weighting parameter $(\alpha)$ on probability calibration, which we explore in the subsequent section. The flexibility of the focal loss function could prove advantageous in future studies, especially when incorporating additional features or diverse types of input data into the model, which may result in more complex classification scenarios.

\subsection{Improving Probability Calibration with Logit-Transformed Platt Scaling}
\label{sec: platt scaling}

\begin{figure*}[ht]
\centering
\includegraphics[scale=.46]{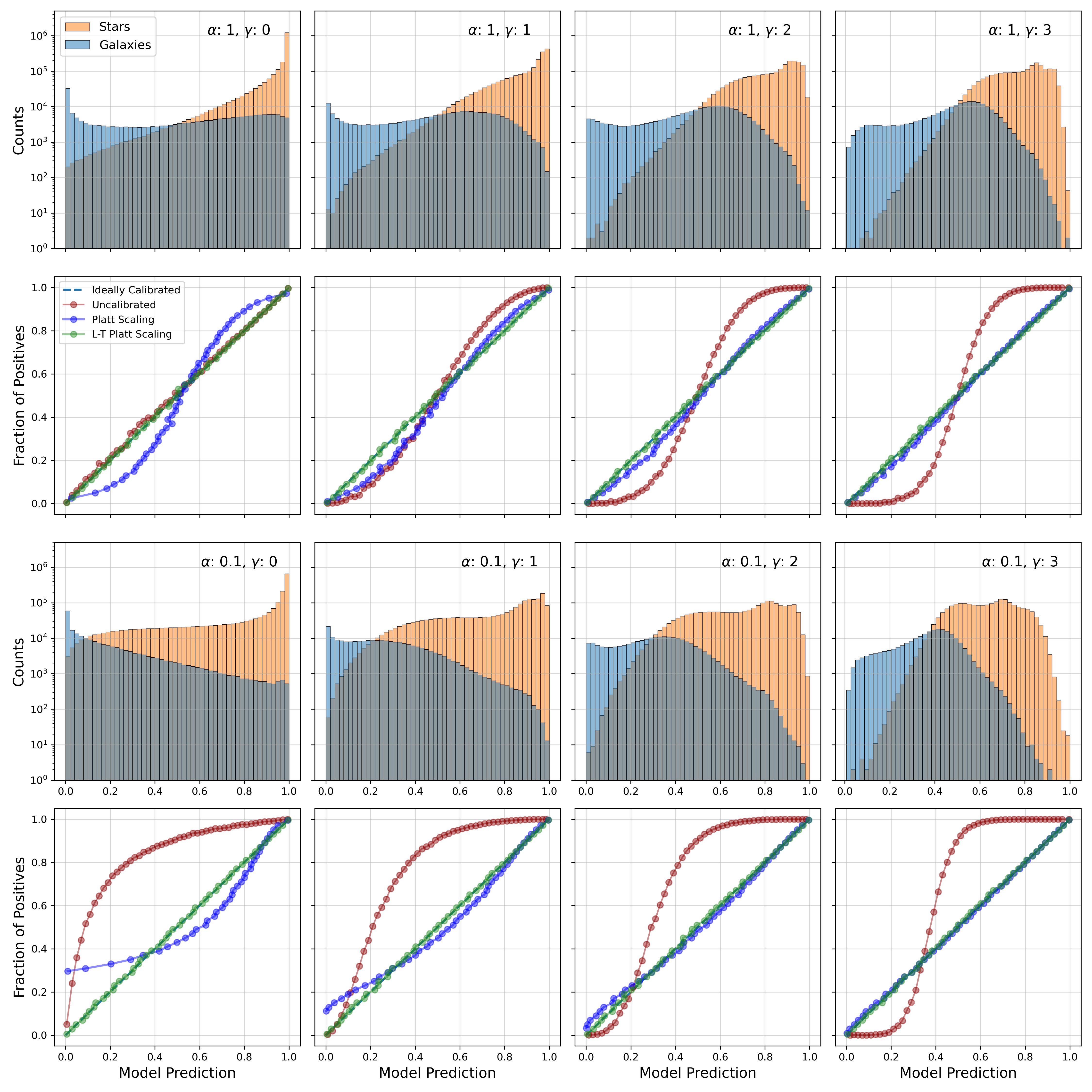}
\caption{Calibration analysis of the output of models with varying $\alpha$ and $\gamma$ focal loss parameters. The first and third rows display the model predictions for the test set, with the colour indicating the true class. The second and fourth rows showcase the calibration results for each corresponding model, presenting the uncalibrated, Platt-scaled, and Logit-Transformed Platt scaling results. This side-by-side comparison illustrates the effectiveness of the calibration techniques in enhancing the alignment of predicted probabilities with the observed frequency of positive class occurrences.}
\label{fig: Calibration}
\end{figure*}

A crucial aspect of a well-performing classifier is its ability to provide reliable posterior class probability estimates that align with the actual likelihood of a positive class occurrence. This characteristic, known as calibration, is often visualized using probability calibration curves or reliability curves \citep{Degroot1983}.

To construct a probability calibration curve, we apply the model to the calibration set, approximately 1 million cutouts, and divide the probability estimates into discrete bins, each representing a range of predicted probabilities. We then calculate the fraction of positive examples for each bin and plot these fractions against the predicted probabilities. A well-calibrated classifier will have a calibration curve closely following the diagonal line, indicating that the predicted probabilities accurately reflect the observed frequency of positive class occurrences. 

However, it is often the case that the raw output scores of a classifier do not perfectly align with these observed frequencies. This discrepancy can be due to various factors, including the complexity of the data, the model's assumptions, and the training process. To address this issue, a variety of post-calibration methods have been developed, aiming to adjust the output scores and improve their alignment with the true probabilities.
One such method is Platt scaling, a widely used technique for calibrating the outputs of a model. Platt scaling transforms the raw output values to better align with the true probabilities of the predicted classes \citep{platt1999}. The standard Platt scaling formula is:

\begin{equation}\label{eq:platt}
P(y=1|f) = \frac{1}{1 + \text{exp}\left(Af + B\right)},
\end{equation}

where $A$ and $B$ are parameters estimated from the data using the Maximum Likelihood method. This method has been shown to improve the calibration and performance of machine learning models \citep{NiculescuMizil2005, kull2019}.
However, Platt scaling can struggle when the output of a classifier is not sigmoid-shaped \citep{kull2017}. To address this, we use a simple modification of the traditional Platt scaling method: applying a logit transformation to the model's output scores prior to Platt scaling. While this transformation is recognized in the calibration literature \citep{Filho2023}, existing studies have not thoroughly investigated its advantages and disadvantages. However, in our specific case, the Logit-Transformed (L-T) Platt scaling shows superior performance, providing empirical evidence of its effectiveness over simple Platt scaling.

The L-T Platt scaling formula is given by:

\begin{equation}\label{eq:logit_platt}
P(y=1|x) = \frac{1}{1 + \text{exp}\left(A \; \text{logit}(f) + B\right) }.
\end{equation}

In the formula above, the term logit(f) in the equation represents the logit transformed output score of the model. The parameters A and B, estimated from the data, are similar to those in the traditional Platt scaling method. These parameters are optimized to achieve the best possible calibration for the specific dataset and model at hand. 

We applied the L-T Platt scaling to a range of models, each trained with different focal loss parameters. The resulting models' predictions and their calibration, illustrated in Fig. \ref{fig: Calibration}, demonstrate that this method can recover almost perfectly calibrated probabilities for each model.

The calibration process is a critical step in our methodology, enabling the direct comparison of different models. As discussed in Section \ref{sec:lossf}, post-calibration, the performance of the models converge, indicating the robustness of our dataset. This robustness obviates the need for correction factors from the focal loss, suggesting that a model trained with a simple cross entropy loss function can directly yield optimal results.

Interestingly, our analysis also suggests that the calibration technique itself may not be necessary for our specific case. However, the process provides valuable insights into the performance of different models and serves as a useful tool for model comparison and evaluation. Therefore, while not essential for our specific task, the calibration process contributes to a more comprehensive understanding of our models' performance and the impact of different loss functions. This knowledge can serve as a valuable guide for future research and model development in star-galaxy classification and beyond. The insights gained can potentially influence a wider range of applications, fostering advancements in the broader field of astronomical image analysis.

\section{Results}
\label{sec: Results}

This section presents a thorough analysis of our network's performance using a test set of over 2 million cutouts. We juxtapose our results with the star-galaxy catalogues generated by MeerLICHT using SourceExtractor, providing a comparative assessment of their respective performances. We employ a dimensionality reduction technique to gain a deeper understanding of the network's learning process. Additionally, we showcase the network's proficiency in classifying sources within crowded fields, particularly those near the Galactic Plane. This comprehensive analysis underscores the robustness and versatility of our network in managing a wide array of challenging astronomical classification scenarios.

We utilise a set of relevant metrics to gauge the effectiveness of our results. Unlike deterministic classifiers that assign discrete labels to each source, our probabilistic classifier provides a probability that determines whether each source is a star or a galaxy. The performance evaluation of probabilistic classifiers often involves transforming probability estimates into class labels by setting a specific probability threshold. For example, a source is classified as a star if $p > 0.5$ and a galaxy if $p \leq 0.5$. However, this approach can be misleading if the predictions are not well-calibrated. Therefore, we employ several performance metrics that are suitable for probabilistic classifiers, the Area Under the Curve (AUC) for the Receiver Operating Characteristic (ROC, \citealp{Swets1996}) curve, the Precision-Recall curve (AUPRC, \citealp{Davis2006}) and Brier score \citep{BrierScore}.

\subsection{Comparison with SourceExtractor}
\label{sec: SourceExtractor}

SourceExtractor \citep{Bertin1996}, a widely-utilized software in the field of astronomy for source extraction, feature estimation, and classification, serves as a benchmark in our study. Despite its extensive use in the astronomical community and its integration into the MeerLICHT official pipeline, SourceExtractor exhibits certain limitations in the context of star-galaxy classification.

\begin{figure*}[ht]
\centering
\includegraphics[scale=0.44]{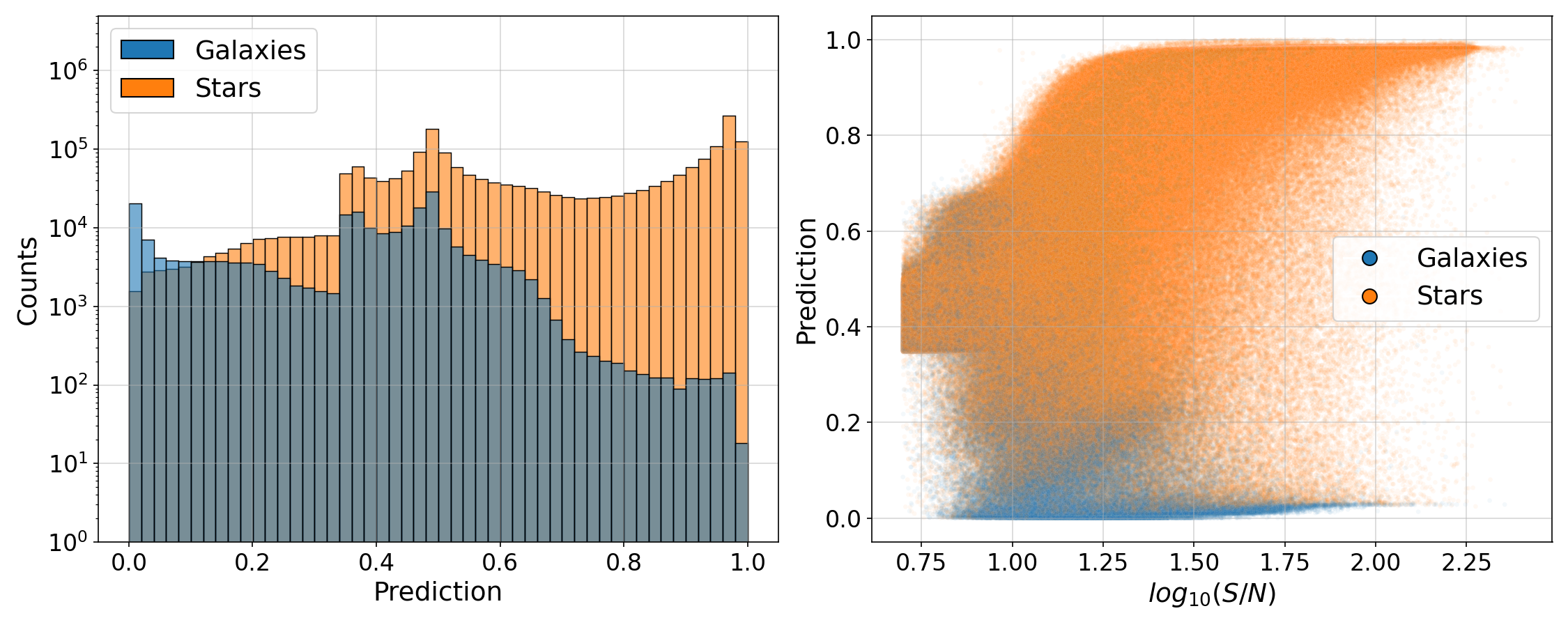}
\caption{Star-galaxy prediction for SourceExtractor. Left: The histogram of the predictions, coloured by true class, shows a bias towards values of $\approx 0.5$, indicating a lack of calibration. Right: Star-Galaxy classification performance of SourceExtractor as a function of source S/N. The figure shows a decrease in classification performance for sources with lower S/N, indicating that SourceExtractor struggles to classify these sources accurately.}
\label{fig: predictions_SE}
\end{figure*}

Fig. \ref{fig: predictions_SE} illustrates the performance of SourceExtractor for star-galaxy classification. The left panel shows the distribution of SourceExtractor's predictions, which displays a bias towards a large number of objects with a probability value of $\approx 0.5$, close to the binary classifier threshold. This bias can increase misclassifications, particularly for galaxies, which are often assigned probabilities close to the 0.5 threshold. The right panel of Fig. \ref{fig: predictions_SE} depicts the performance of SourceExtractor as a function of S/N. It becomes evident that the tool struggles with sources with a lower S/N, leading to decreased classification performance for these sources.

\begin{figure*}[ht]
\centering
\includegraphics[scale=0.44]{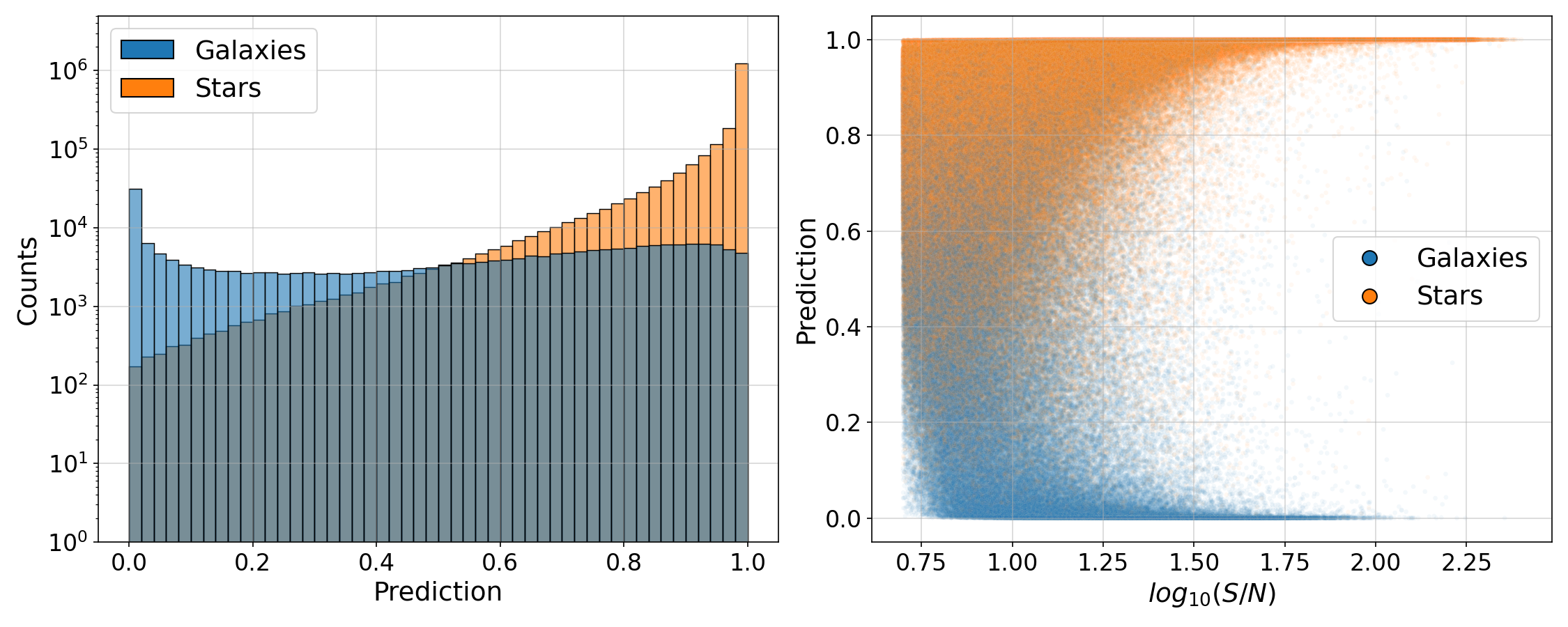}
\caption{Star-galaxy prediction for ASID-C. Left: The histogram of the predictions, coloured by true class, shows a well-calibrated set of predictions. Right: Star-Galaxy classification performance of ASID-C as a function of source S/N. The figure shows that ASID-C maintains a high level of performance across a wider range of S/N values, thereby outperforming SourceExtractor, particularly for sources with low to medium S/N.}
\label{fig: predictions_ASID_C}
\end{figure*}

In contrast, our method, ASID-C, generates a more refined and well-calibrated prediction probability compared to SourceExtractor, as depicted in Fig. \ref{fig: predictions_ASID_C}. The left panel shows the distribution of ASID-C's star-galaxy predictions, which are more evenly distributed and well-calibrated. The right panel of Fig. \ref{fig: predictions_ASID_C} demonstrates superior performance of ASID-C, particularly for sources with low to medium S/N. ASID-C is more robust and effective in classifying sources across a wider range of S/N values, thereby outperforming SourceExtractor.

The comparative analyses presented in this section underscore the effectiveness of ASID-C in star-galaxy classification, particularly in challenging scenarios involving sources with lower S/N. To further substantiate this, we evaluate the performance of the classifiers using the metrics introduced earlier: ROC-AUC, PR-AUC, and Brier score, calculated as a function of source S/N.
The ROC-AUC and PR-AUC metrics assess the balance between sensitivity and specificity, and the precision-recall trade-off in the classifier's predictions, respectively. The Brier score, on the other hand, quantifies the accuracy of probabilistic predictions by measuring the mean squared difference between the predicted probabilities assigned to the possible outcomes and the actual outcome. A lower Brier score indicates more accurate predictions, with a score of 0 representing a perfect classifier.

Table \ref{tab:PerformanceMetrics} provides a comprehensive view of the performance metrics across different S/N values for ASID-C, SourceExtractor (SE), and a baseline model (Base), where the labels are randomly assigned with a 10:1 proportion. The best performance for each metric and S/N value is highlighted in bold font. This table allows for a detailed comparison of the classifiers' performance, further substantiating the superior effectiveness of ASID-C in star-galaxy classification.

\begin{table}[ht]
\caption{Performance Metrics by S/N. The AUPRC, due to its formulation, is calculated separately for both stars (s) and galaxies (g) as the positive class. The table shows that ASID-C consistently outperforms both SE and the baseline model across most S/N values and metrics.}
\label{tab:PerformanceMetrics}
\begin{tabular}{c|ccccc}
\hline
\hline
\noalign{\smallskip}
S/N & 0.84 & 1.12 & 1.40 & 1.69 & 1.98 \\
\noalign{\smallskip}
\hline
\hline
\noalign{\smallskip}
ROCAUC ASID & \textbf{0.837} & \textbf{0.937} & \textbf{0.974} & \textbf{0.983} & 0.981 \\
ROCAUC SE & 0.581 & 0.850 & 0.968 & 0.981 & \textbf{0.985} \\
ROCAUC Base & 0.500 & 0.500 & 0.501 & 0.498 & 0.499 \\
\hline
\noalign{\smallskip}
AUPRC ASID (s) & \textbf{0.952} & \textbf{0.990} & \textbf{0.998} & \textbf{1.000} & \textbf{1.000} \\
AUPRC SE (s) & 0.841 & 0.975 & \textbf{0.998} & \textbf{1.000} & \textbf{1.000} \\
AUPRC Base (s) & 0.813 & 0.893 & 0.950 & 0.984 & 0.997 \\
\hline
\noalign{\smallskip}
AUPRC ASID (g) & \textbf{0.609} & \textbf{0.781} & \textbf{0.895} & \textbf{0.941} & \textbf{0.946} \\
AUPRC SE (g) & 0.271 & 0.575 & 0.782 & 0.853 & 0.782 \\
AUPRC Base (g) & 0.187 & 0.107 & 0.050 & 0.016 & 0.003 \\
\hline
\noalign{\smallskip}
Brier ASID & \textbf{0.108} & \textbf{0.045} & \textbf{0.012} & \textbf{0.002} & \textbf{0.000} \\
Brier SE & 0.279 & 0.188 & 0.076 & 0.034 & 0.015 \\
Brier Base & 0.250 & 0.185 & 0.140 & 0.113 & 0.101 \\
\noalign{\smallskip}
\hline
\end{tabular}
\end{table}

From the table, it is evident that ASID-C consistently outperforms both SE and Base across most S/N values and metrics. ASID-C particularly excels in low S/N regions, where classifying stars and galaxies becomes more challenging. This superior performance is reflected in the higher ROC-AUC and PR-AUC values and lower Brier scores for ASID-C compared to SE and Base. The significantly lower Brier scores for ASID-C suggest that it generates more reliable and well-calibrated probability predictions, especially for galaxies near the detectability threshold.

In the context of our study, the high AUPRC values observed for all three methods, particularly for stars as the positive class, can be attributed to the imbalance in the dataset, which is heavily skewed towards stars. We evaluated the AUPRC with both stars and galaxies as the positive class to gain a comprehensive understanding of the classifier's performance across both classes. This approach is particularly important in our case due to the significant class imbalance.
The AUPRC is a metric that is particularly informative in the context of imbalanced datasets. However, a high AUPRC does not necessarily imply that the classifier is performing well. Precision, a component of the AUPRC, is calculated as TP / (TP + FP) and can remain high even if the number of False Positives (FP) increases as long as the number of True Positives (TP) is much larger. Therefore, while the AUPRC values indicate that ASID-C, SE, and the baseline model are able to classify a large number of stars correctly, it is crucial also to consider the full set of performance metrics of Table \ref{tab:PerformanceMetrics}.

Given the significant classification challenges posed by low S/N regions, the ability of ASID-C to handle them effectively not only underscores its robustness but also its potential as a highly effective tool for star-galaxy classification in future astronomical studies.

\subsection{Model performance with UMAP}
\label{sec: UMAP}

To gain a deeper understanding of how ASID-C differentiates between stars and galaxies, we employ Uniform Manifold Approximation and Projection (UMAP \footnote{\url{https://umap-learn.readthedocs.io/en/latest/index.html}}, \citealp{McInnes2018}), a dimensionality reduction technique. UMAP proposed as an alternative to T-distributed Stochastic Neighbor Embedding (t-SNE) \citep{t-sne}, is renowned for its ability to preserve the global structure of data and its efficient implementation.

We first adapted the trained model to integrate UMAP into our deep neural network by removing the final classification layer (of the model depicted in Fig. \ref{fig: architecture}). This adjustment enables us to concentrate on the output from the preceding 32-dimensional dense layer, which is abundant with insights about the features learned by the network.
Subsequently, we employed this adapted model to generate predictions on the test set and apply UMAP to the latter. UMAP provides a condensed summary of the decision-making process, unveiling the intricate patterns the model utilizes to make accurate classifications and offering a more nuanced understanding of the network's performance. This method highlights the complex interplay of features contributing to the successful differentiation between stars and galaxies.

\begin{figure}[ht]
\centering
\includegraphics[scale=0.34]{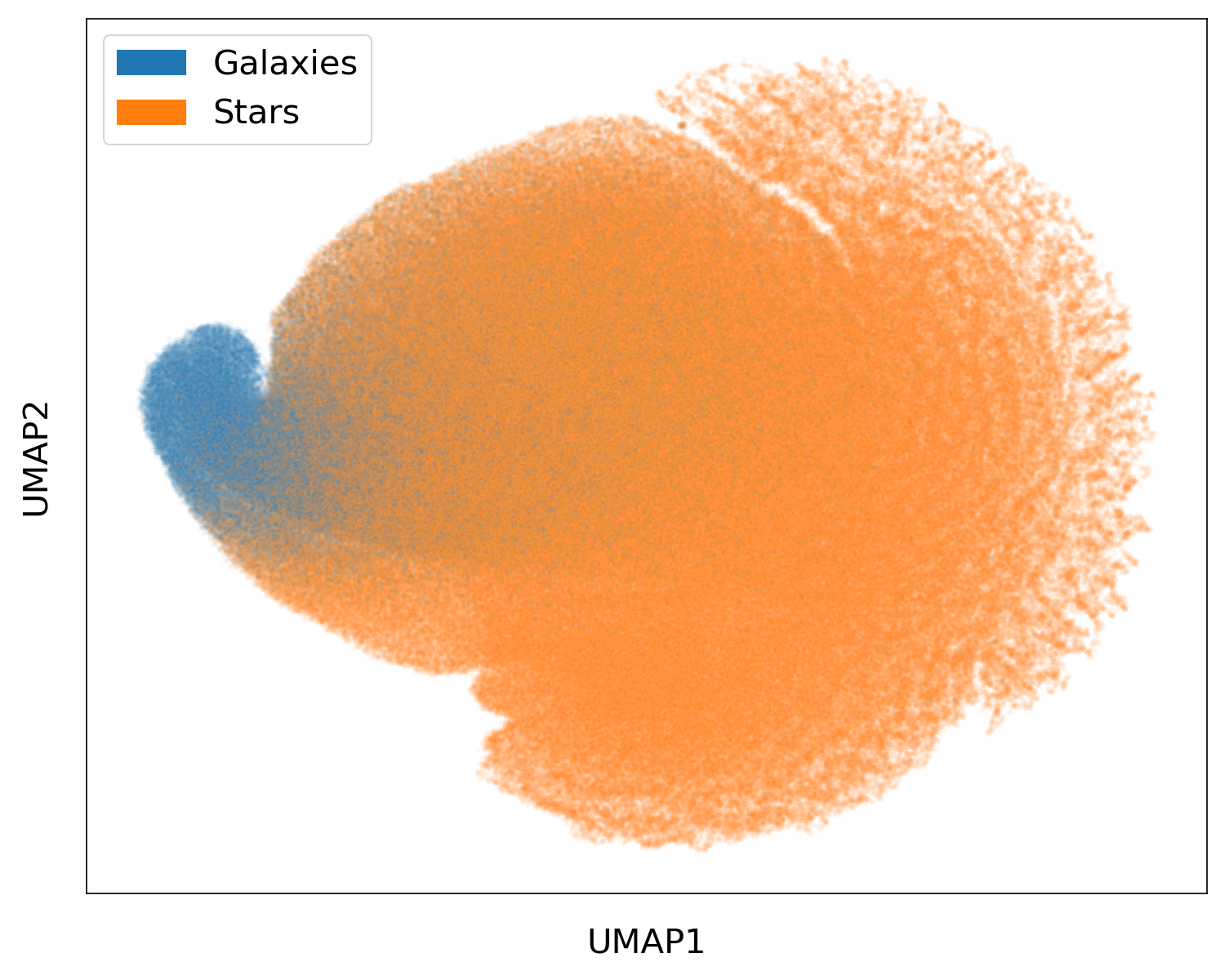}
\caption{UMAP latent space visualization. The axes represent the UMAP embeddings, with each point coloured by their label, star (orange) or galaxy (blue).}
\label{fig: UMAP}
\end{figure}

As depicted in Fig. \ref{fig: UMAP}, the UMAP visualization reveals a distinct separation between stars and galaxies, with a clear clump corresponding to galaxies. The intermediate region, which upon further checks, includes sources with a low S/N, emphasizes the complexity inherent in distinguishing between these two classes.
It is important to note that the axes in Fig. \ref{fig: UMAP} represent the UMAP embeddings and do not lend themselves to direct interpretation as with techniques like Principal Component Analysis (PCA). The construction of a UMAP visualization requires selecting specific parameters that will influence the final output. In our case, we chose a Correlation metric, 30 nearest neighbours, and a minimum distance of 1.

\begin{figure}[ht]
\centering
\includegraphics[scale=0.31]{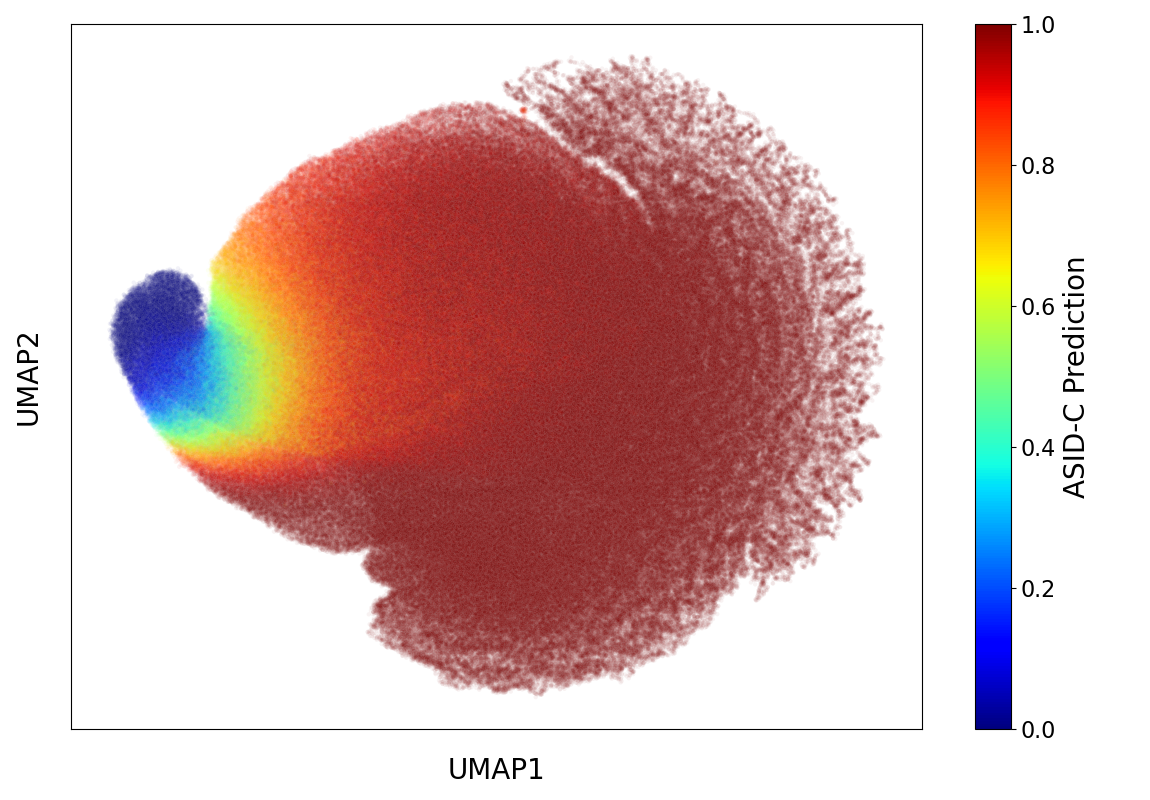}
\caption{UMAP latent space visualization. The axes represent the UMAP embeddings, with each point coloured by their full-model predicted value.}
\label{fig: UMAP_predictions}
\end{figure}

Fig. \ref{fig: UMAP_predictions} presents the UMAP embeddings coloured by their full model prediction. This visualization further emphasizes the separation between galaxies, predominantly predicted with a 0, and stars, predominantly predicted with a 1. The intermediate region of predictions, ranging between 0.4 and 0.6, corresponds to the area where distinguishing between the two classes is most challenging.

The insights gleaned from this visualization proved invaluable during the training process, facilitating refinements in model design and contributing to our understanding of the problem. By visualizing the decision-making process of our model, we can better comprehend its strengths and limitations and use this understanding to enhance its performance in future applications.

\subsection{ASID-C Performance in High Stellar Density Regions}
\label{sec: Crowded}

In this section, we showcase the applicability of ASID-C on images oriented towards the Galactic Plane, a region teeming with stars and with a typical number of galaxies that are, however, difficult to differentiate. These areas, distinguished by a markedly high stellar density and complex background, traditionally present considerable challenges for classification techniques. Factors such as significant interstellar extinction, confusion with Galactic structures, and high source overlap further complicate the task. Despite the relative scarcity of galaxies compared to stars, their accurate identification and classification are crucial for various astronomical studies, particularly for identifying transients' host galaxies and understanding the structure of our own Galaxy. 

For the analysis presented in this section, we applied a threshold to the predictions, considering sources with a predicted value less than 0.5 as galaxies. Although not optimal, this threshold was chosen based on the distribution of predictions and the known characteristics of the dataset.

\begin{figure*}[ht]
\centering
\includegraphics[scale=0.36]{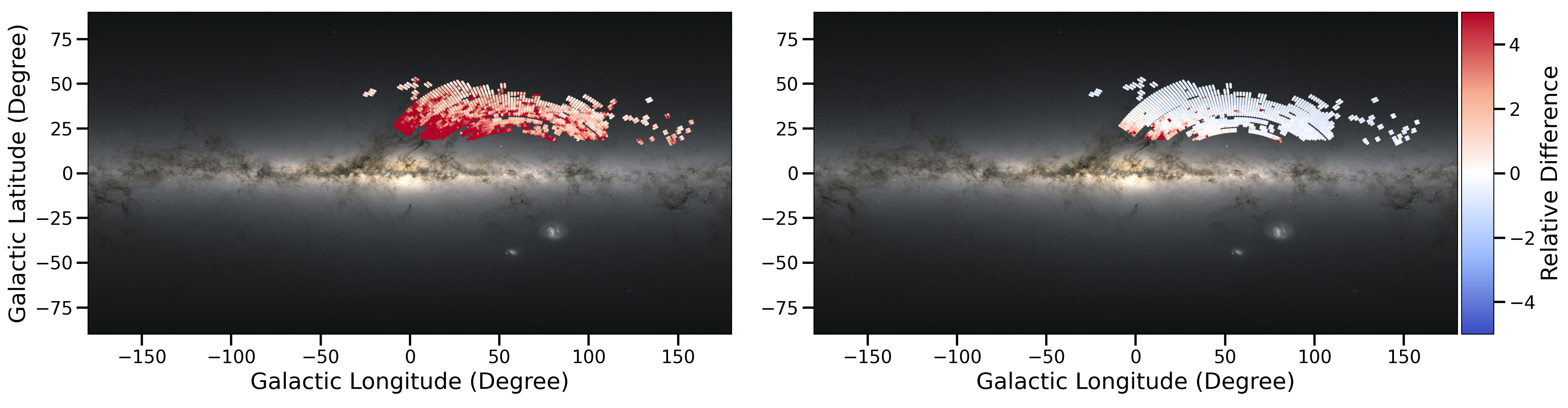}
\caption{Comparison of SourceExtractor (left) and ASID-C (right) performance across the Galactic Plane. The tiles are coloured based on the relative difference between the actual and estimated numbers of galaxies in that region. The left panel highlights the tendency of SourceExtractor to overestimate galaxies in regions with high stellar density, indicating a bias in its classification approach. The right panel showcases the ability of ASID-C to classify galaxies accurately, even in regions with high stellar density and complex structures, thereby enabling reliable identification of celestial objects in such challenging environments.}
\label{fig: crowded_field_comparison}
\end{figure*}

Fig. \ref{fig: crowded_field_comparison} illustrates the performance of both SourceExtractor (left panel) and ASID-C (right panel) across the Galactic Plane. The tiles are coloured based on the relative difference between the actual and estimated numbers of galaxies in that region.
The left panel of Fig. \ref{fig: crowded_field_comparison} shows the performance of SourceExtractor, which tends to overestimate galaxies in regions with high stellar density, indicating a bias in its classification approach.
In contrast, the right panel of Fig. \ref{fig: crowded_field_comparison} showcases the performance of ASID-C in the same regions. ASID-C accurately classifies galaxies almost independently of the number of sources in the region. Even amidst the high density of stars and complex structures close to the Galactic Plane, ASID-C enables reliable identification of galaxies.

The robust performance of ASID-C in high stellar density regions demonstrates its versatility and potential for broad applications in astronomy. Its ability to accurately classify galaxies amidst complex structures and dense star populations can significantly enhance our understanding of celestial phenomena.

\subsection{Evaluation of Prediction Timings}
\label{sec: PredictionTimings}

The computational efficiency of a model is a critical factor, especially when dealing with large astronomical datasets. To evaluate the computational efficiency of ASID-C, we measured the time taken to predict the class of different sizes of datasets on a GPU-accelerated system, which significantly benefits deep learning models like ASID-C.
The datasets used in this analysis ranged from 5,000 to 100,000 images. We performed the prediction fifty times for each dataset size and calculated the average time taken. This approach mitigates potential variability in the timing results due to factors such as system load or GPU thermal throttling.

\begin{figure}[ht]
\centering
\includegraphics[scale=0.34]{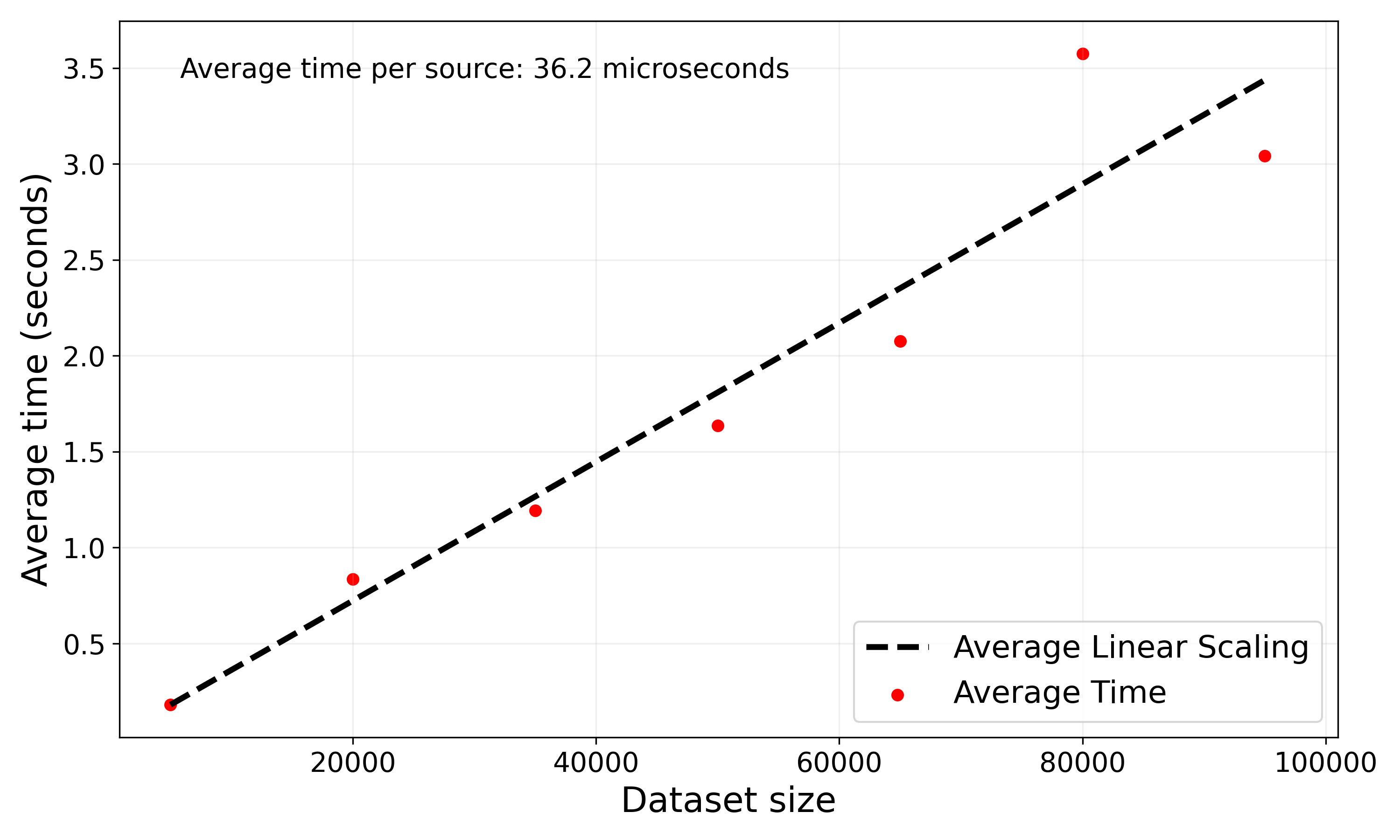}
\caption{Average time taken by ASID-C to predict on different sizes of datasets. The red dots represent the average time taken per dataset size, and the black dashed line represents the expected time based on linear scaling.}
\label{fig: Timings}
\end{figure}

The results, shown in Fig. \ref{fig: Timings}, demonstrate that the time ASID-C took to predict the class of the images scales linearly with the dataset size. This linear scaling is a desirable property, as it means that the time required to process a large dataset can be accurately estimated based on the time taken to process a smaller dataset sample.
ASID-C demonstrates excellent computational efficiency, with linear scaling of prediction time with dataset size and a time per source of approximately 36 microseconds. This efficiency, combined with its predictions' high accuracy and reliability, makes ASID-C a powerful tool for star-galaxy classification in large astronomical surveys.

\section{Conclusion and Discussion}
\label{sec: Conclusion and Discussion}

This study introduced ASID-C, a dual-branch convolutional neural network (CNN) specifically designed for star-galaxy classification. ASID-C incorporates both image data and spatial information, specifically the location of the source within the image, to provide a robust and effective solution for distinguishing between stars and galaxies in astronomical images from the MeerLICHT telescope.

Our evaluation of ASID-C's performance demonstrates its superiority over the widely-used SourceExtractor tool, particularly regarding the reliability and calibration of its probability predictions. Notably, ASID-C exhibits exceptional performance for low signal-to-noise (S/N) sources, a challenging area where many existing tools struggle. This enhanced accuracy for low S/N sources significantly contributes to the robustness of source classification and analysis, thereby enhancing the quality and reliability of astronomical research.

We further tested ASID-C's performance in dense source regions close to the Galactic Plane. The results underscore the model's ability to maintain high accuracy even in crowded regions, which are traditionally challenging for classification tasks. This capability enables researchers to accurately classify celestial objects in these regions, providing valuable insights into the formation and evolution of stars and galaxies, as well as the dynamics and structure of the Galactic Plane.

In addition to its classification performance, we explored the latent space of the trained ASID-C model using the UMAP technique. This analysis revealed the underlying structures and patterns within the data that facilitate the distinction between stars and galaxies, providing further insights into the model's classification process.

A standout feature of ASID-C's performance is the near-perfect calibration of its predictions, particularly for models trained with cross-entropy loss. While Platt scaling, especially when enhanced with a logit transformation, can improve the calibration of model predictions, we found that the cross-entropy models already yield almost perfectly calibrated predictions. This ensures that the predicted probabilities closely mirror the actual class proportions, a critical attribute for many applications. A well-calibrated model can significantly improve decision-making processes, particularly when an accurate estimate of class membership probability is paramount.

One of the most promising applications of ASID-C lies in real-time star-galaxy classification. The rapid processing speed of convolutional neural networks, combined with the robustness of ASID-C, makes it an ideal tool for time-sensitive astronomical observations. For instance, transient events, such as supernovae or gamma-ray bursts, require immediate follow-up observations to capture their rapidly changing properties. By providing an accurate and immediate classification of sources, ASID-C can help astronomers quickly identify the nature of the transient's host galaxy or rule out false positives, thereby streamlining the transient detection and follow-up process.

Beyond binary star-galaxy classification, ASID-C could potentially be extended to classify galaxies based on their morphology. Galaxy morphology, such as spiral, elliptical, or irregular, provides valuable insights into the formation and evolution of galaxies. By training ASID-C on a dataset labelled with galaxy morphologies, we could leverage its robust feature extraction capabilities to identify the morphological characteristics of galaxies. This would expand the capabilities of ASID-C and contribute to our understanding of galaxy evolution.

While our current work focuses on optical data, ASID-C has the potential to handle data across multiple wavelengths, such as infrared, ultraviolet, or radio. This multi-wavelength approach could enhance classification accuracy by providing a more comprehensive view of the sources. For instance, certain types of galaxies or stars may exhibit unique characteristics at specific wavelengths that are not apparent in optical data. By incorporating multi-wavelength data into ASID-C, we can leverage these unique characteristics to improve classification performance.

In conclusion, ASID-C represents a significant advancement in the field of star-galaxy classification. Its strong performance and adaptability make it a valuable tool for the astronomical community, opening up new opportunities for research and study. By enhancing our ability to classify celestial objects accurately, ASID-C contributes to our understanding of the universe and accelerates the pace of discovery and exploration in astronomy. The insights gained from this study provide valuable guidance for future research and applications in star-galaxy classification.

\begin{acknowledgements}
F.S. and G.N. acknowledge support from the Dutch Science Foundation NWO. S.B. and G.Z. acknowledge the financial support from the Slovenian Research Agency (grants P1-0031, I0-0033 and J1-1700). R.R. acknowledges support from the Ministerio de Ciencia e Innovación (PID2020-113644GB-I00). G.P. acknowledges support by ICSC – Centro Nazionale di Ricerca in High Performance Computing, Big Data and Quantum Computing, funded by European Union – NextGenerationEU.
The MeerLICHT telescope is a collaboration between Radboud University, the University of Cape Town, the South African Astronomical Observatory, the University of Oxford, the University of Manchester and the University of Amsterdam, and supported by the NWO and NRF Funding agencies.
The authors would like to thank Dr Dmitry Malyshev for his valuable suggestions and insightful discussions, which contributed to improving this paper.\\

The Legacy Surveys consist of three individual and complementary projects: the Dark Energy Camera Legacy Survey (DECaLS; Proposal ID \#2014B-0404; PIs: David Schlegel and Arjun Dey), the Beijing-Arizona Sky Survey (BASS; NOAO Prop. ID \#2015A-0801; PIs: Zhou Xu and Xiaohui Fan), and the Mayall z-band Legacy Survey (MzLS; Prop. ID \#2016A-0453; PI: Arjun Dey). DECaLS, BASS and MzLS together include data obtained, respectively, at the Blanco telescope, Cerro Tololo Inter-American Observatory, NSF’s NOIRLab; the Bok telescope, Steward Observatory, University of Arizona; and the Mayall telescope, Kitt Peak National Observatory, NOIRLab. Pipeline processing and analyses of the data were supported by NOIRLab and the Lawrence Berkeley National Laboratory (LBNL). The Legacy Surveys project is honored to be permitted to conduct astronomical research on Iolkam Du’ag (Kitt Peak), a mountain with particular significance to the Tohono O’odham Nation.

NOIRLab is operated by the Association of Universities for Research in Astronomy (AURA) under a cooperative agreement with the National Science Foundation. LBNL is managed by the Regents of the University of California under contract to the U.S. Department of Energy.

This project used data obtained with the Dark Energy Camera (DECam), which was constructed by the Dark Energy Survey (DES) collaboration. Funding for the DES Projects has been provided by the U.S. Department of Energy, the U.S. National Science Foundation, the Ministry of Science and Education of Spain, the Science and Technology Facilities Council of the United Kingdom, the Higher Education Funding Council for England, the National Center for Supercomputing Applications at the University of Illinois at Urbana-Champaign, the Kavli Institute of Cosmological Physics at the University of Chicago, Center for Cosmology and Astro-Particle Physics at the Ohio State University, the Mitchell Institute for Fundamental Physics and Astronomy at Texas A\&M University, Financiadora de Estudos e Projetos, Fundacao Carlos Chagas Filho de Amparo, Financiadora de Estudos e Projetos, Fundacao Carlos Chagas Filho de Amparo a Pesquisa do Estado do Rio de Janeiro, Conselho Nacional de Desenvolvimento Cientifico e Tecnologico and the Ministerio da Ciencia, Tecnologia e Inovacao, the Deutsche Forschungsgemeinschaft and the Collaborating Institutions in the Dark Energy Survey. The Collaborating Institutions are Argonne National Laboratory, the University of California at Santa Cruz, the University of Cambridge, Centro de Investigaciones Energeticas, Medioambientales y Tecnologicas-Madrid, the University of Chicago, University College London, the DES-Brazil Consortium, the University of Edinburgh, the Eidgenossische Technische Hochschule (ETH) Zurich, Fermi National Accelerator Laboratory, the University of Illinois at Urbana-Champaign, the Institut de Ciencies de l’Espai (IEEC/CSIC), the Institut de Fisica d’Altes Energies, Lawrence Berkeley National Laboratory, the Ludwig Maximilians Universitat Munchen and the associated Excellence Cluster Universe, the University of Michigan, NSF’s NOIRLab, the University of Nottingham, the Ohio State University, the University of Pennsylvania, the University of Portsmouth, SLAC National Accelerator Laboratory, Stanford University, the University of Sussex, and Texas A\&M University.

BASS is a key project of the Telescope Access Program (TAP), which has been funded by the National Astronomical Observatories of China, the Chinese Academy of Sciences (the Strategic Priority Research Program “The Emergence of Cosmological Structures” Grant \# XDB09000000), and the Special Fund for Astronomy from the Ministry of Finance. The External Cooperation Program of Chinese Academy of Sciences (Grant \# 114A11KYSB20160057) and Chinese National Natural Science Foundation (Grant \# 12120101003, \# 11433005) also support the BASS.

The Legacy Survey team makes use of data products from the Near-Earth Object Wide-field Infrared Survey Explorer (NEOWISE), which is a project of the Jet Propulsion Laboratory/California Institute of Technology. NEOWISE is funded by the National Aeronautics and Space Administration.

The Legacy Surveys imaging of the DESI footprint is supported by the Director, Office of Science, Office of High Energy Physics of the U.S. Department of Energy under Contract No. DE-AC02-05CH1123, by the National Energy Research Scientific Computing Center, a DOE Office of Science User Facility under the same contract; and by the U.S. National Science Foundation, Division of Astronomical Sciences under Contract No. AST-0950945 to NOAO.

\end{acknowledgements}

\bibliography{Bibliography}

\begin{appendix}

\section{Incorporating Spatial Information}
\label{sec: appendix_I}

The dual-branch structure of our model, ASID-C, is designed to incorporate a variety of features. A key component we've integrated is spatial information, which plays a crucial role in astronomical image analysis. The appearance of celestial objects can vary depending on their location in the image, complicating the task of distinguishing between stars and galaxies.
To address this, ASID-C considers the location of each celestial object within the image. This approach allows the model to account for variations in the Point Spread Function (PSF) and other spatially-dependent effects, enhancing its accuracy and reliability even under challenging imaging conditions.

\begin{figure}[h]
\centering
\includegraphics[scale=0.39]{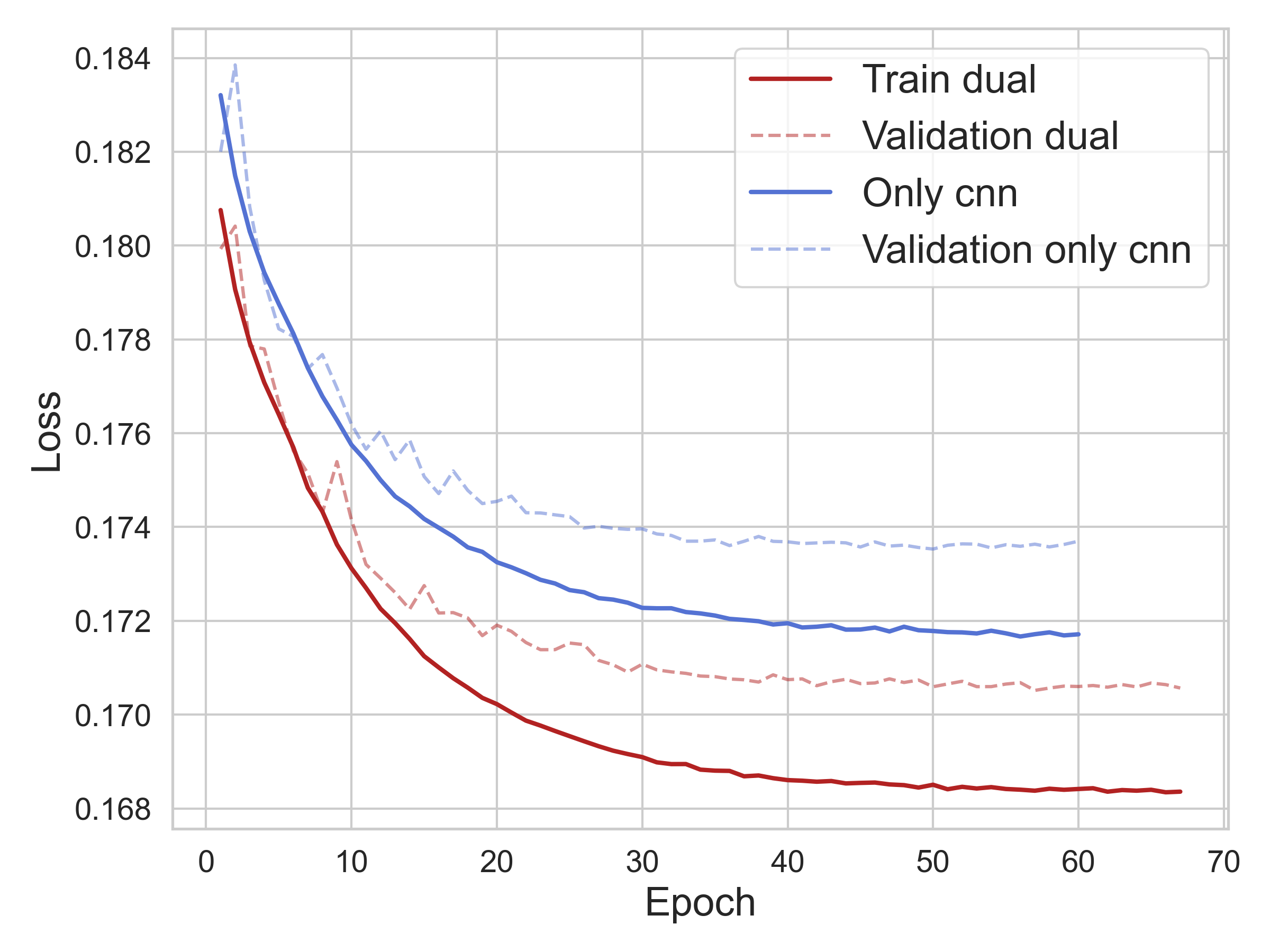}
\caption{Comparison of loss between models with and without the spatial information branch. The model incorporating spatial information exhibits a lower loss, indicating improved performance.}
\label{fig: loss_comparison}
\end{figure}

\begin{figure}[H]
\centering
\includegraphics[scale=0.39]{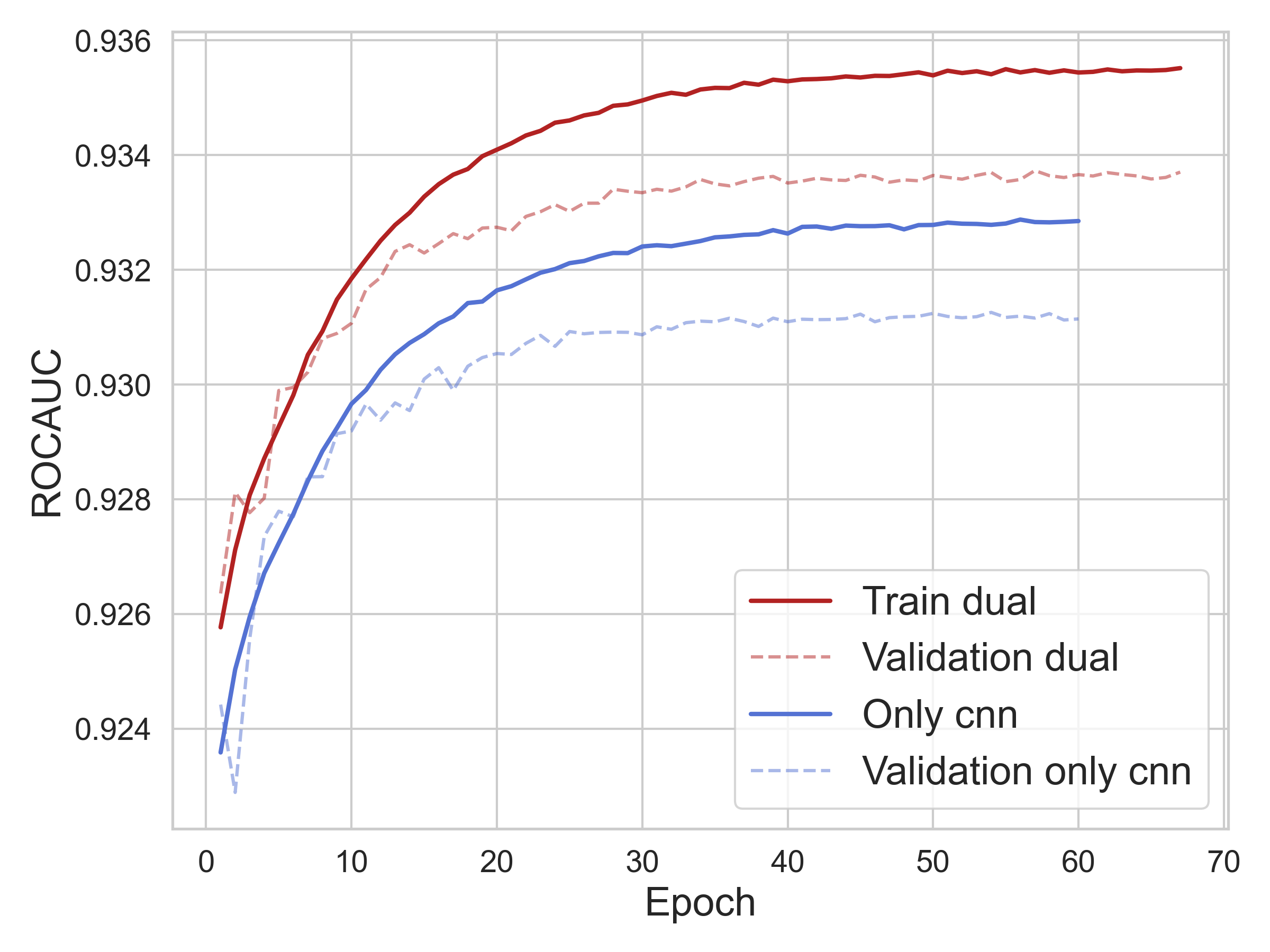}
\caption{Comparison of AUC between models with and without the spatial information branch. The model incorporating spatial information achieves a higher AUC, demonstrating its superior classification performance.}
\label{fig: AUC_comparison}
\end{figure}

\begin{figure}[H]
\centering
\includegraphics[scale=0.39]{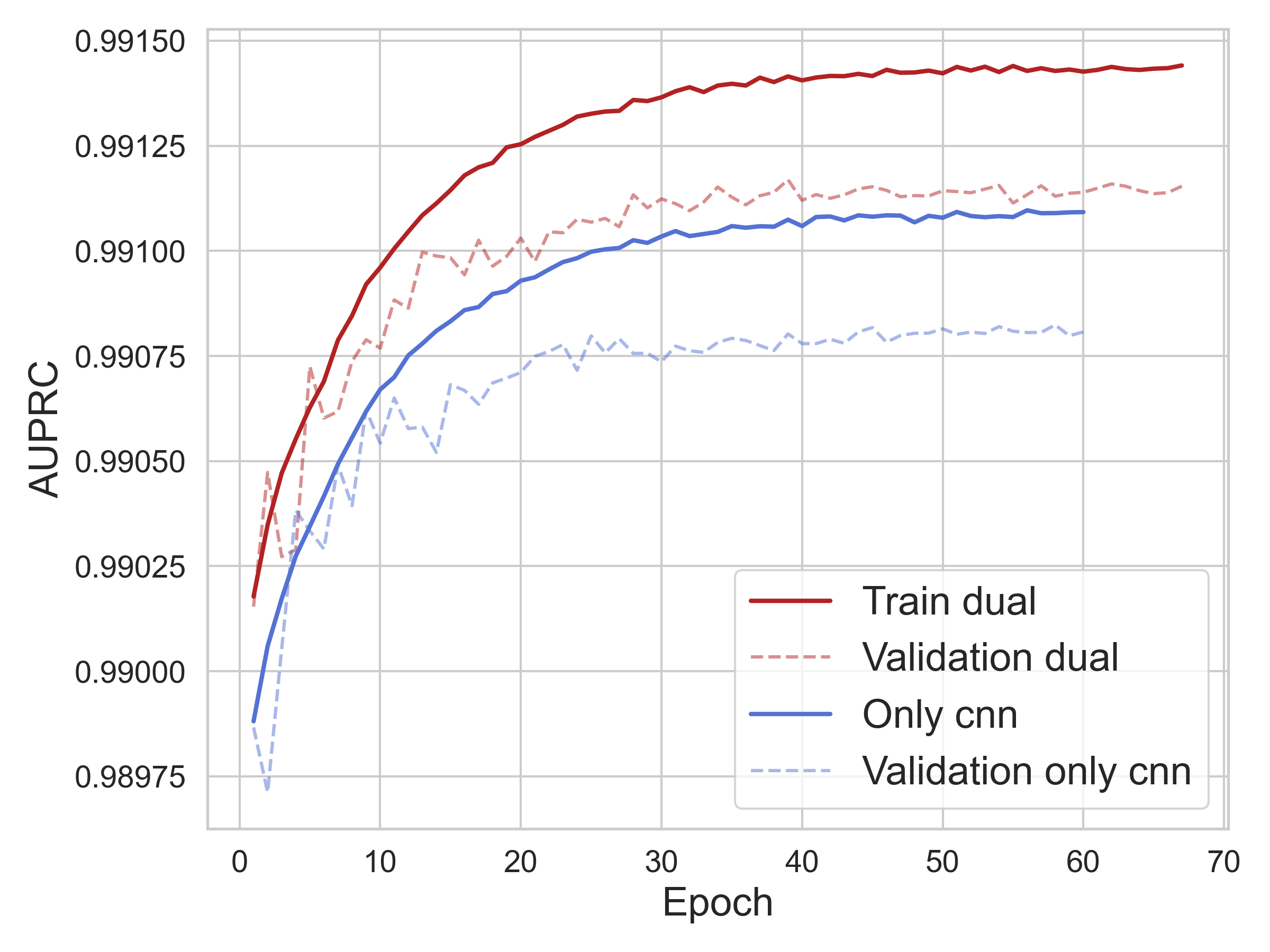}
\caption{Comparison of AUPRC between models with and without the spatial information branch. The model incorporating spatial information shows a higher AUPRC, indicating a better balance between precision and recall.}
\label{fig: AUPRC_comparison}
\end{figure}

To validate the effectiveness of this approach, we compared the performance of the model with and without the spatial information branch. The results showed a significant improvement when spatial information was included, underscoring its importance in star-galaxy classification.
Figures \ref{fig: loss_comparison}, \ref{fig: AUC_comparison}, and \ref{fig: AUPRC_comparison} illustrate the comparison of loss, Area Under the Receiver Operating Characteristic Curve (AUC), and Area Under the Precision-Recall Curve (AUPRC) respectively, between the models with and without the spatial information branch.

Moreover, the dual-branch structure of ASID-C allows for the inclusion of other features, such as PSF information, colour data from multi-band images, or metadata from external catalogues. This adaptability ensures our model can evolve to meet the changing needs and challenges of star-galaxy classification.

\end{appendix}

\end{document}